\documentclass[aps,prc,reprint,showpacs,groupedaddress,onecolumn]{revtex4-1}

\usepackage{graphicx,color} 
\usepackage{dcolumn}  
\usepackage{bm}       
\usepackage{amsmath}
\usepackage{xcolor}

\begin{document}

\newcommand {\nc} {\newcommand}
\nc {\IR} [1]{\textcolor{red}{#1}}
\nc {\IB} [1]{\textcolor{blue}{#1}}
\nc {\IP} [1]{\textcolor{violet}{#1}}
\nc {\IG} [1]{\textcolor{olive}{#1}}
\nc {\IT} [1]{\textcolor{teal}{#1}}

\title{Quantifying uncertainties due  to irreducible three-body forces in 
deuteron-nucleus reactions}

\author{L. Hlophe} 
\email{hlophe1@llnl.gov}
\affiliation{Lawrence Livermore National Laboratory, P.O. Box 808, L-414, Livermore, CA 94551, USA}

\author{K. Kravvaris}
\affiliation{Lawrence Livermore National Laboratory, P.O. Box 808, L-414, Livermore, CA 94551, USA}

\author{S. Quaglioni}
\affiliation{Lawrence Livermore National Laboratory, P.O. Box 808, L-414, Livermore, CA 94551, USA}

\date{\today}

\begin{abstract} 
\noindent{\bf Background:}
 While Faddeev techniques enable the exact description of the three-body dynamics, their predictive power is limited in part by the omission of irreducible neutron-proton-nucleus  three-body force ($n$-$p$-$A$ 3BF).
\\
{\bf Purpose:}
  Our goal is to quantify systematic uncertainties due to the omission of the 3BF arising from the full antisymmetrization of the ($A$+$2$)-body system with the $\alpha$ particle fixed in its ground state, using using as testing grounds $d$+$\alpha$ scattering and the $^6$Li ground state.      
 \\ 
{\bf Methods:}
We adopt the ab initio no-core shell model coupled with the resonating group method (NCSM/RGM) to compute microscopic $n$-$\alpha $ and $p$-$\alpha$ interactions, and use them in a three-body description of the $d$+$\alpha$ system by means of momentum-space Faddeev-type equations. Simultaneously, we also carry out ab initio calculations of $d$+$\alpha$ scattering and $^6$Li ground state by means of six-body NCSM/RGM calculations to serve as a benchmark for the three-body model predictions given by the Faddeev calcualtions.
\\
{\bf Results:}
By comparing the Faddeev and NCSM/RGM results, we show that the irreducible $n$-$p$-$\alpha$ 3BF has a non-negligible effect on bound state and scattering observables alike. Specifically, the Faddeev approach 
yields a $^6$Li ground state that is approximately $600$~keV shallower than the one obtained  with the NCSM/RGM. Additionally, the Faddeev calculations for $d$+$\alpha$ scattering yield a $3^+$ resonance that is located approximately $400$~keV higher in energy compared to  the NCSM/RGM result. The shape of the 
$d$+$\alpha$ angular distributions computed using the two approaches also differ, owing to the discrepancy in the predictions of the $3^+$ resonance energy.
\\
{\bf Conclusions:}
The Faddeev three-body model predictions for $d$+$\alpha$ scattering and $^6$Li using microscopic $n$-$\alpha$ and $p$-$\alpha$ potentials differ from those computed microscopically with the NCSM/RGM. These discrepancies are due to the $n$-$p$-$\alpha$ 3BF which arises from two-nucleon exchange terms in the microscopic $d$-$\alpha$ interaction and are not accounted for in the three-body model Faddeev calculations. This study lays the foundation for future parametrizations of the 3BF due to Pauli exclusion Principle effects in improved three-body calculations of deuteron-induced reactions.
\end{abstract}


\maketitle

\section{Introduction}
\label{intro}
Deuteron-induced nuclear  reactions are powerful tools for probing the properties of nuclei, especially for short-lived exotic isotopes for which direct measurements are unfeasible. Additionally, combining experimental measurements with an accurate reaction theory provides means for determining nuclear structure information as well as capture rates that are needed for astrophysical nucleosysnthesis modeling. An ab initio description of such deuteron-induced reactions requires the solution of the many-body scattering problem, which at this time is only feasible for light systems with up to nine nucleons~\cite{Hupin:2018biv,Raimondi:2016njp}. To overcome this limitation, deuteron-induced scattering and reactions on a target nucleus ($A$) are typically  treated within a three-body model consisting of a
dynamically inert core, a proton ($p$), and a neutron ($n$) interacting through an effective three-body
Hamiltonian, such that an exact solution can be obtained through the Faddeev
formalism~\cite{Faddeev:1961}.  While simplified numerical approaches have been applied with varying degrees of
success~\cite{rcjohnson:1970,Johnson:1974uh,Yahiro:1986}, the use of the Faddeev formalism eliminates additional errors due to approximate treatment of three-body dynamics. 

Typically, the effective three-body Hamiltonian
is approximated as a direct sum of the pairwise interactions of the subsystems, namely, the  $n$-$p$, $n$-$A$, and $p$-$A$ potentials. The $n$-$p$ potential is usually constrained 
by high-precision 
fits to nucleon-nucleon (NN) elastic scattering data. Commonly adopted models are, for example, the Bonn~\cite{Machleidt:1987hj,Machleidt:2000ge} or 
chiral effective theory~\cite{Entem:2003ft,Entem:2014msa,Epelbaum:2014sza} ($\chi$EFT)
potentials. Local phenomenological nucleon-nucleus (${\cal W}_{NA}$) potentials constrained using elastic
scattering data have been successfully applied to describe deuteron-induced reactions on targets ranging from
the lightest nuclei, e.g., the deuteron~\cite{Deltuva:2006ch} to much heavier nuclei such as
nickel~\cite{Deltuva:2009zc}. However, such nucleon-nucleus ($N$-$A$) potentials are not uniquely defined and,
as a result, different functional forms provide accurate fits to $N$+$A$ scattering data but yield different
predictions for the deuteron-nucleus ($d$+$A$) systems~\cite{Lei:2018toi,Titus:2014rfa,Doleschall:2008zz}.
Further, insights from microscopic reaction theory revealed that the nucleon-nucleus potentials are generally
non-local and dispersive~\cite{Feshbach:1958wf}. The work of Ref.~\cite{Titus:2014rfa} demonstrated that
non-locality significantly impacts predictions for $d$+$A$ reaction observables. Moreover, it was shown in Ref.~\cite{Charity:2007zza} that imposing dispersivity leads to improved nucleon-nucleus potentials that
consistently describe both the bound and scattering states for the $N$+$A$ system.
Finally, the formal projection of the ($A$+$2$)-body problem onto the three-particle space produces not only
pairwise potentials but also an irreducible effective three-body force (3BF). To improve the predictive power of the three-body model it is thus necessary to investigate the origin of this 3BF and
assess its impact on properties of the $d$+$A$ system. Previous investigations along these lines include the
work of Ref.~\cite{Dinmore:2019ouu}, where the 3BF emerges from interactions between the neutron and proton via excitation of the target $A$.

In the present work we investigate the irreducible 3BF in a light nuclear system and quantify its effects on observables using as a test case the $d$+$\alpha$ system. Our objective is to isolate contributions to the 3BF arising from the Pauli exclusion
principle, disentangling them from other mechanisms such as target excitation and three-nucleon (3N) forces. Starting from a five-nucleon Hamiltonian based on the $\chi$EFT NN interaction of Ref.~\cite{Entem:2003ft,Entem:2014msa} softened with a similarity
renormalization group (SRG) transformation~\cite{Bogner:2007} (in two-body space), momentum-space non-local microscopic $n$-$\alpha$ (${\cal W}_{n\alpha}$ ) and
$p$-$\alpha$ (${\cal W}_{p\alpha}$ ) potentials are computed working within the framework of the no-core shell model (NCSM) coupled with the resonating group method (RGM)~\cite{Quaglioni:2008sm,Quaglioni:2009mn}. 
For the purpose of the present study,
we restrict the four nucleons within the $\alpha$-particle to the ground state.
These non-local, momentum-space ${\cal W}_{N\alpha}$ potentials are then used together with the SRG-evolved $\chi$EFT $n$-$p$ interaction within the Faddeev formalism to make predictions for the $d$+$\alpha$ system. To overcome the limitations of the screening and renormalization approach introduced in Ref.~\cite{deltuva:06b},
which is numerically stable only in regions where the effects of the $p$+$A$ Coulomb potential are relatively weak (i.e., at high energies and in light nuclei), here we utilize a new capability~\cite{hlophe:2022} based on
the Alt-Grassberger-Sandhas (AGS)~\citep{AGS} formulation of the Faddeev equations in a basis of $p$+$A$
Coulomb scattering wavefunctions~\cite{Mukhamedzhanov:2012qv}.
Simultaneously, we also apply the (six-body) NCSM/RGM approach to directly compute $d$+$\alpha$ bound and scattering observables 
starting from the same $\chi$EFT NN interaction. By comparing the Faddeev and NCM/RGM results, we show that the irreducible 3BF owing to microscopic antisymmetrization effects has a small but significant impact on the
$^6$Li bound state and $d$+$\alpha$ scattering, accounting for $600$ keV of the binding energy and shifting the $3^+$ resonance by $\sim400$ keV lower in energy.

This paper is organized as follows. In Section~\ref{formal:3b} we present a brief summary of the Faddeev-AGS equations in the Coulomb basis. An overview of the momentum space NCSM/RGM needed to evaluate the effective
$N$-$A$ potentials is provided in Sec.~\ref{formal:ai} and a detailed discussion with explicit expressions for
the potentials is given in Appendix~\ref{appendixA}. 
In Sect.~\ref{discuss}, we discuss our results for $d$+$\alpha$ scattering and $^6$Li bound-state calculations, and
assess the effect of the irreducible three-body force by comparing results obtained within the Faddeev-AGS
approach with those computed directly with the NCSM/RGM. The conclusion and outlook are given in Section~\ref{conclusion}.
\section{Formalism}
\label{sec:formalism}
\label{formal}
\subsection{Faddeev formalism with exact treatment of the average Coulomb potential}
\label{formal:3b}

\begin{figure}[!ht]
\begin{center}
\includegraphics[width=14cm]{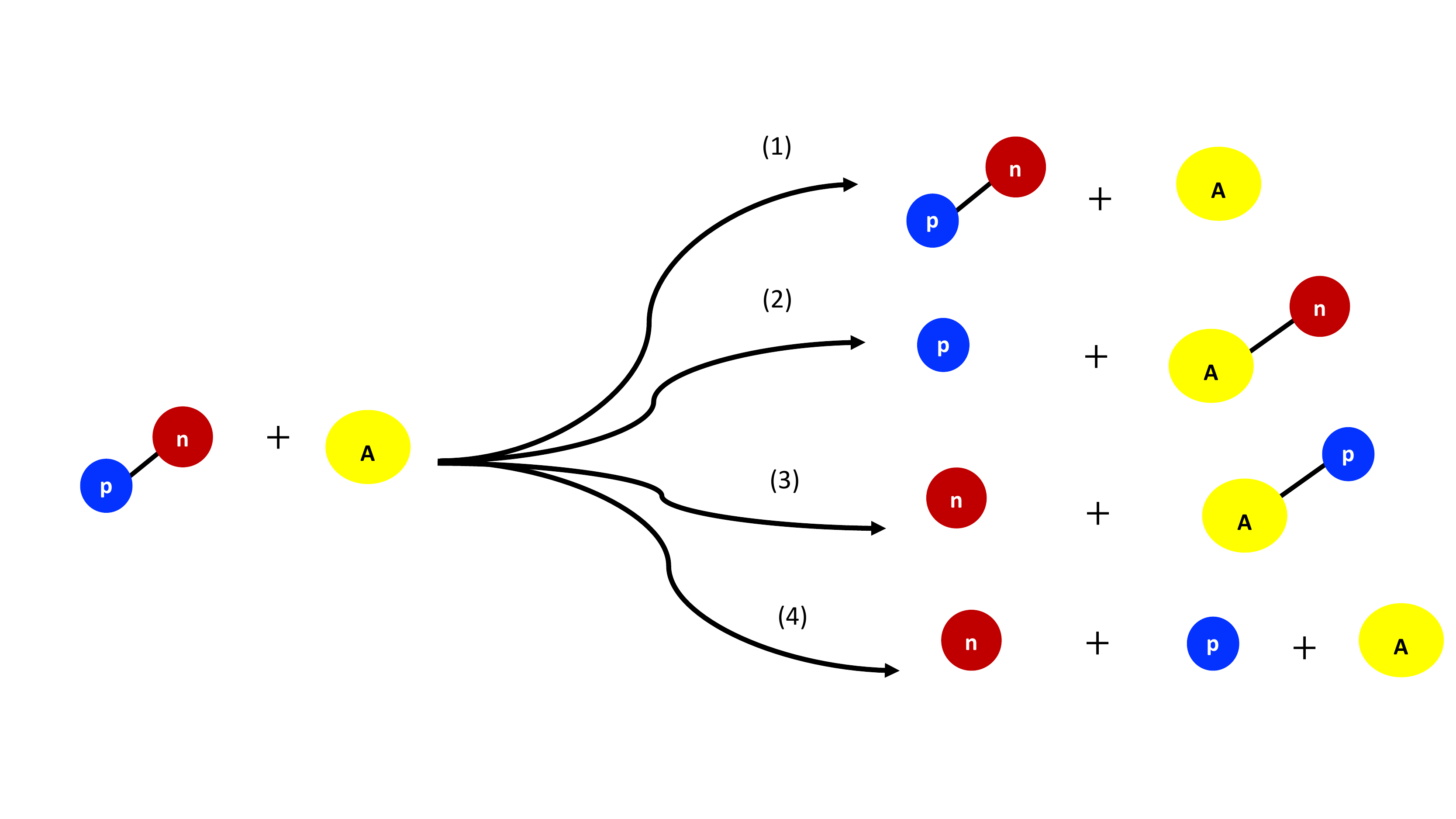}
\vspace{3mm}
\caption{Schematic representation of all possible three-body rearrangement processes arising from the scattering of a deuteron with a target nucleus $A$. The arrows (1) - (4) represent elastic scattering, neutron transfer, proton transfer, and three-body breakup.      
}
\label{fig1}
\end{center}
\end{figure}

The scattering of  a deuteron from a target nucleus $A $ leading to all possible three-body rearrangement processes (see Fig.~\ref{fig1}) can be consistently described using the Faddeev formalism. It is convenient to
assign numerical labels $\{1,\;2,\;3\}$ to the particles $\{A,\;p,\;n\}$, respectively, and to define
corresponding arrangement channels $i=1,2,3$ consisting of a spectator particle $i$ and the remaining particle
pair interacting through their respective pairwise potential ${\cal W}_{i}$. For example, arrangement channel
$1$ contains an interacting $n+p$ pair with a spectator $A$, where ${\cal W}_{1}\equiv V^{np}$. Similarly,
${\cal W}_{2}\equiv {\cal W}^{nA}$, and ${\cal W}_{3}\equiv {\cal W}^{pA}$. The binary potentials generally have a non-local dependence on the momentum coordinates $p$ and $p'$ or ${\cal W}^{N-A}={\cal W}^{N-A,\;
{\cal I}^\pi}_{\nu\nu'}(p,p')$, where $\nu$ $(\nu')$ represents angular momentum channels and ${\cal I}^\pi$ is
the spin-parity of the system. For each arrangement channel $i$,
in the center-of-mass (c.m.) frame, the system can be represented by a pair of Jacobi momenta: the relative
momentum of the pair ($\vec{p}_i\equiv p_i\;\hat{p_i}$) and the momentum of the spectator with respect to the c.m. of the pair ($\vec{q}_i\equiv q_i\;\hat{q_i}$).
As such, the kinetic energy is given by~\cite{Hlophe:2017bkd} 
\begin{eqnarray}
H_0=\frac{p_{i}^2}{2\mu_{i}}+\frac{q_i^2}{2M_i},
\label{eq:2.1}
\end{eqnarray}
where $\mu_i$ and $M_i$ are respectively the reduced masses for the interacting pair ($jk$) and for the $i$+$(jk)$ system, where ($i,j,k$) form a cyclic permutation of ($1,2,3$). At the relative three-body energy $E$, the exact
three-body scattering wave function for the incident arrangement channel $1$  ($|\Psi_{(1)}\rangle$) fulfills
the Schr\"odinger equation
\begin{eqnarray}
\left[E-H_0\right]|\Psi_{(1)}\rangle=\sum\limits_{i=1}^3 {\cal W}_i|\Psi_{(1)} \rangle.
\label{eq:2.2}
\end{eqnarray}
The asymptotics of the wave function $|\Psi_{(1)} \rangle$ contain information about all possible three-body processes and therefore has complicated boundary conditions that make its direct determination unfeasible. To circumvent this issue, Faddeev~\citep{Faddeev:1961}
introduced the components $|\psi_{i1} \rangle \equiv G_0(E)\;{\cal W}_i|\Psi_{(1)} \rangle$ so that the wave
function is given by the sum $|\Psi_{(1)} \rangle = |\psi_{11} \rangle + |\psi_{21} \rangle +|\psi_{31}
\rangle$, where the free propagator has the definition $G_0(E) = [E-H_0+i0]^{-1}$.  
Unlike the full wave function, the asymptotics of each Faddeev component describe a specific reaction channel, in the present case of deuteron-induced reactions $|\psi_{11} \rangle$, $|\psi_{21} \rangle$, and $|\psi_{31} \rangle$
describe elastic deuteron scattering, neutron transfer, and proton transfer reactions, respectively. 

Since scattering wave functions are ill-behaved in momentum space, it is customary to work with transition operators instead. In the case of three-body scattering the transition operators $U^{11}$, $U^{21}$, and
$U^{31}$, contain the same information as the corresponding Faddeev components and fulfill a set of coupled
momentum space integral equations~\cite{AGS} (Faddeev-AGS equations)
\begin{eqnarray}
U^{i1}(E)=\bar\delta_{i1}\;G_0^{-1}(E)+\sum\limits_{k=1}^{3}\bar\delta_{ik}\;t_{k}(E)\;G_0(E)\;U^{k1}(E).
\label{eq:2.3}
\end{eqnarray}
Here, $\bar\delta_{ik}\equiv 1-\delta_{ik}$ is the anti-Kronecker delta and $t_i(E)$ is the binary $t$~matrix given by the Lippmann-Schwinger (LS) equation $t_i(E)={\cal W}_i+{\cal W}_i\;G_0(E)\;t_i(E)$. For processes
leading to complete three-body breakup the corresponding transition operator is given by $U^{01}=U^{11}+U^{31}+U^{31}$. The transition amplitudes are connected to the scattering matrix ($s$~matrix)
through the relation $S^{i1}=\delta_{i1}-2 \pi i \delta(E-H_0)\;U^{i1} $. We proceed by introducing the states $\left|q_i
\left({\lambda_i I_i^{\pi_i}}\right){\cal J}_i M_{{\cal J}_i}\right\rangle$, which describe the motion of the
spectator relative to the pair with $\lambda_i$ and $I_i^{\pi_i}$ representing the orbital angular momentum and
spin-parity of the spectator, respectively. Here ${\cal J}_i$ and $M_{{\cal J}_i}$ denote the total spectator
angular momentum and its $z$-axis projection. Additionally,
the bound state wave function of the pair $\left(\left|\phi_{im}^{{\cal I}_i^{\pi_i}}\right\rangle \right)$ is
given by the Schr\"odinger equation
\begin{eqnarray}
\left[\varepsilon_i^{(m)}-\frac{p_i^2}{2\mu_i}\right]\left|\phi_{im}^{{\cal I}_i^{\pi_i}}\right\rangle={\cal W}_i^{{\cal I}_i^{\pi_i}}\left|\phi_{im}^{{\cal I}_i^{\pi_i}}\right\rangle,
\label{eq:2.4}
\end{eqnarray}
where $m= 1\;...\;N_{\rm bound}$ is an index enumerating the bound states and $N_{\rm bound}$ is the number of bound states. Here $\varepsilon_i^{(m)}$ is the binding energy of the $m^{th}$ bound state and ${\cal I}^{\pi_i}$ is the spin-parity of the pair.
The channel state corresponding to an arrangement channel $i$ and a conserved total angular momentum $J$ is constructed by coupling the bound state of the pair to the relative motion of spectator leading to
\begin{equation}
\left|\Phi_{m\alpha_i}^{(i),JM}\right\rangle\left| q_i\right\rangle \equiv \left|q_i\right\rangle \left[\left|\phi_{im}^{{\cal I}_i^{\pi_i}}\right\rangle\left| (\lambda_i I_i){\cal J}_i\right\rangle\right]^{JM},
\label{eq:2.5}
\end{equation}
where $\alpha_i=\{{\cal I}^{\pi_i},\lambda_i,I_i^{\pi_i},{\cal J}_i\}$ and $M$ is the $z$-axis projection of $J$. The transition matrix elements needed to evaluate the cross section for the process $1\longrightarrow i$ are then given by 
\begin{eqnarray}
X^{i1,J}_{m\alpha_i,n\alpha_i'}(q_i,q_1';E)&\equiv&\left\langle q_i\right|\left\langle \Phi_{m\alpha_i}^{(i),JM}\Big|\;U^{i1}(E)\;\Big|\Phi_{n\alpha_1}^{(1),JM}\right\rangle\left| q_1'\right\rangle,\cr\cr
&\equiv&\left\langle q_i\right|X^{i1,J}_{m\alpha_i,n\alpha_1'}(E) \left| q_1'\right\rangle.
\label{eq:2.6}
\end{eqnarray}
For cross section calculations we only need transition matrix elements for which 
$q_{im}$ and $q_{1n}'$ are determined by the on-shell condition.

To obtain numerical solutions of the Faddeev-AGS equations, we first note that the pairwise potentials have the general form ${\cal W}_{i}={\cal W}_i^{\rm s}+\bar V_i^{\rm c}$, where ${\cal W}_i^{\rm s}$ is short-ranged and $\bar
V_i^{\rm c}=Z_jZ_k e^2/r$ is the average Coulomb potential, with $Z_j (Z_k)$ being the charge of particle $j$($k$). The corresponding binary $t$~matrix can be written as $t_i(E)=t_i^{\rm c}(E)+t^{\rm s}_i(E)$, where
$t_i^{\rm c}(E)$  is the $t$~matrix associated with the average Coulomb potential and $t_i^{\rm s}$ is the solution of the LS equation using the short-ranged residual interaction ${\cal W}_i^{\rm s}\equiv {\cal W}_i-\bar V^{\rm c}_i$ in
the basis of Coulomb scattering wave functions. The presence of the term $t_i^{\rm c}(E)$ introduces non-integrable singularities in the kernel of Eqs.~(\ref{eq:2.3}) and thus renders a direct numerical solution
unfeasible~\cite{AGS,Mukhamedzhanov:2000qg}. By subtracting  $t_i^{\rm c}(E)$ from the overall $t$~matrix, one can reformulate the Faddeev-AGS equations leading to~\citep{AGS}
 \begin{eqnarray}
U^{i1}(E)=U^{{\rm c},i1}(E)+\sum\limits_{k=1}^{3}U^{{\rm c},ik}(E)\;t_{k}^{\rm s}(E)\;G_0(E)\;U^{k1}(E), 
 \label{eq:2.7}
 \end{eqnarray}
 where $U^{{\rm c},ik}$ is the solution of the Faddeev-AGS equations with only the average Coulomb potential and has an analytical solution when only two of the three particles are charged. To arrive at a numerical solution for Eqs~(\ref{eq:2.7}), we express the short-ranged pairwise potentials in separable form via expansion in a basis $\{|\bar h^i_n\rangle\}$ (e.g. Ernst-Shakin-Thaler~\cite{Ernst:1973zzb} basis),
 \begin{eqnarray}
 {\cal W}_i^{\rm s}(E)&= \sum\limits_{\beta,\gamma=1}^{N_{\rm rank}} | \bar h^i_{\beta}\rangle\;\lambda_{i,\beta\gamma}^{\rm s}\;\langle \bar h^{i}_{\gamma}|,
\label{eq:2.8}
 \end{eqnarray}
where $N_{\rm rank}$ is the number of basis functions which are enumerated by the indices $\beta(\gamma)$ hereafter. This representation for ${\cal W}_i^{\rm s}$ leads to a corresponding separable form of the binary $t$~matrix 
 \begin{eqnarray}
 t_i^{\rm s}(E)&= \sum\limits_{n=1}^{N_{\rm bound}}|\tilde h^i_{n}\rangle\;g_{in}^{0}(E)\;\langle \tilde h^{i}_{n}| +\sum\limits_{\beta\gamma=1}^{N_{\rm rank}} |\bar h^i_{\beta}\rangle\;\bar\tau_{i,\beta\gamma}^{\rm s}(E)\;\langle \bar h^{i}_{\gamma}|,
\label{eq:2.9}
 \end{eqnarray}
 where $g_{in}^{0}(E)=(E-\varepsilon_i^{(n)}-q_i^2/M_i+i0)^{-1}$ is the free propagator for the spectator relative to the bound pair. The vectors $|\tilde h^i_{n}\rangle$ appearing in the first term of Eq.~(\ref{eq:2.9}) are the so-called bound state form factors and are given by the product of the pairwise potential and the bound state
 wave function or $|\tilde h^i_{n}\rangle \equiv {\cal W}_i|\phi_{m}^{(i)}\rangle$. The matrix elements $\bar\tau_{i,\beta\gamma}^{\rm s}(E)$ are determined by substituting Eqs.~(\ref{eq:2.8}) and (\ref{eq:2.9}) into
 the LS equation. 
 The separable representation of the binary $t$~matrix enables a reformulation of Eqs.~(\ref{eq:2.7}) in terms of the transition operators defined by Eq.~(\ref{eq:2.6}) such that~\cite{AGS} 
 \begin{eqnarray}
 X^{ij}_{mn}(E)=Z^{ij}_{mn}(E)&+&\sum\limits_{k=1}^3\sum\limits_{n'=1}^{N_{\rm bound}}\; Z^{ik}_{mn'}(E)\;g_{kn'}^{0}(E)\;X^{kj}_{n'n}(E)\cr\cr
&+& \sum\limits_{k=1}^3\sum\limits_{\beta',\gamma'=1}^{N_{\rm rank}}\; Z^{ik}_{m\beta'}(E)\;\bar\tau^{\rm s}_{k,\beta'\gamma'}(E)\;X^{kj}_{\gamma' n}(E),
\label{eq:2.10a}
\end{eqnarray}
 and
  \begin{eqnarray}
 X^{ij}_{\beta n}(E)=Z^{ij}_{\beta n}(E)&+&\sum\limits_{k=1}^3\sum\limits_{n'=1}^{N_{\rm bound}}\; Z^{ik}_{ \beta n'}(E)\;g_{kn'}^{0}(E)\;X^{kj}_{n'n}(E)\cr\cr
&+& \sum\limits_{k=1}^3\sum\limits_{\beta',\gamma'=1}^{N_{\rm rank}}\; Z^{ik}_{\beta \beta'}(E)\;\bar\tau^{\rm s}_{k,\beta'\gamma'}(E)\;X^{kj}_{\gamma' n}(E),
\label{eq:2.10b}
\end{eqnarray}
 where we have suppressed the channel index $\alpha_i$ for brevity and the second equation is needed to determine the non-physical transition operators $X^{kj}_{\gamma'n}(E)$. The effective potentials have the definition
 \begin{eqnarray}
 Z^{ij}_{nm}(E)\equiv \left\langle \tilde h^i_{m}\right|G_0(E)\;U^{{\rm c},ij}(E)\;G_0(E)\left|\tilde h^j_{n}\right\rangle,
 \label{eq:zpotij}
 \end{eqnarray}
 and their computation is generally very complicated since $U^{{\rm c},ij}(E)$ is the transition operator for the scattering of three particles under the exclusive influence of the average Coulomb potential. However, it simplifies immensely for cases where one of the particles is neutral. For example, if particle 3 is neutral the Coulomb three-particle transition operator reduces to $U^{{\rm c },ij}(E)= \bar\delta_{ij}\; G_0^{-1}(E)+\bar\delta_{i3}\;t_3^{\rm c}(E)$, where
 $t_3^{\rm c}(E)$ is the two-body Coulomb scattering $t$~matrix for particles 2 and 3. The effective potentials can be numerically evaluated using
  \begin{equation}
  Z^{ij}_{mn}(E) \equiv \bar\delta_{ij}\;\langle h^i_m| \;G_0(E)\;|h^j_n\rangle +\bar\delta_{i3}\;\bar\delta_{j3}\;\langle h^i_m|\;G_0(E)\; t_3^{\rm c}(E)\;G_0(E)\;|h^j_n\rangle.
 \label{eq:2.11}
\end{equation} 
It was already shown in Ref.~\cite{Mukhamedzhanov:2000nt} that the momentum space matrix elements of the diagonal effective potentials $\langle \vec{q_i}'|Z^{ii}_{mn}|\vec{q_i}\rangle$ contain a singularity in the forward direction ($\vec{q_i}\longrightarrow \vec{q_i}'$). The nature of the singularity is revealed by
replacing the Coulomb $t$~matrix with its Born approximation $t_3^{\rm c} = \bar V_{3}^{\rm c}$, leading to the
factorization of the matrix elements $\langle \vec{q_i}'|Z^{ii}_{mn}|\vec{q_i}\rangle$ $=\bar\delta_{i3}\;\langle h^i_m|\;G_0(E,q_i')\;G_0(E,q_i)\;|h^i_n\rangle$ $\times \bar V_{3}^{\rm c}(|\vec{q_i}'-\vec{q_i}|)$.
The forward singularity of the average Coulomb potential is thus propagated into the diagonal effective
potentials, leading to non-integrable singularities in the kernels of Eqs.~(\ref{eq:2.10a}) and (\ref{eq:2.10b}). To circumvent this
challenge, the Born term is subtracted and treated analytically according to the
Gell-Mann-Goldberger~\cite{GellMann:1953zz} relation (two-potential formula). Specifically, one defines the regularized
effective potentials 
 \begin{equation}
  Z^{{\rm sc},ij}_{mn} =Z^{ij}_{mn}-\bar{V}^{\rm c}_3\;\delta_{ij}\;\delta_{mn},
 \label{eq:2.12}
\end{equation} 
so that the full transition operator is given by the sum $X^{ij}_{mn}=X^{{\rm c},ii}_{mm}\;\delta_{mn}+\Omega_{im}^{{\rm c}}\;X^{{\rm sc},ij}_{mn}\;\Omega_{jn}^{{\rm c}}$, where $X^{{\rm c},ii}_{mm}$ corresponds to the
Rutherford scattering amplitude,  $\Omega_{im}^{{\rm c}}(E)\equiv g_{im}^{\rm c}(E)\;g_{im}^{0,-1}(E)$ is the M\"oller wave operator with $g_{im}^{\rm c}(E)=(E-\varepsilon_i^{(m)}-q_i^2/M_i-\bar V^{\rm c}_i+i0)^{-1}$ being the Coulomb propagator. The regularized transition operators $X^{{\rm sc},ij}_{mn}$ are free of singularities and are obtained by solving the modified Faddeev-AGS equations~\citep{AGS},
  \begin{eqnarray}
 X^{{\rm sc},ij}_{mn}(E)=Z^{{\rm sc},ij}_{mn}(E)&+&\sum\limits_{k=1}^3\sum\limits_{n'=1}^{N_{\rm bound}}\; Z^{{\rm sc},ik}_{mn'}(E)\;g_{kn'}^{\rm c}(E)\;X^{{\rm sc},kj}_{n'n}(E)\cr\cr
&+& \sum\limits_{k=1}^3\sum\limits_{\beta',\gamma'=1}^{N_{\rm rank}}\; Z^{{\rm sc},ik}_{m \beta'}(E)\;\bar\tau^{{\rm sc}}_{k,\beta'\gamma'}(E)\;X^{{\rm sc},kj}_{\gamma'n}(E),
\label{eq:2.13a}
\end{eqnarray}
and
  \begin{eqnarray}
 X^{{\rm sc},ij}_{\beta n}(E)=Z^{{\rm sc},ij}_{\beta n}(E)&+&\sum\limits_{k=1}^3\sum\limits_{n'=1}^{N_{\rm bound}}\; Z^{{\rm sc},ik}_{\beta n'}(E)\;g_{kn'}^{\rm c}(E)\;X^{{\rm sc},kj}_{n'n}(E)\cr\cr
&+& \sum\limits_{k=1}^3\sum\limits_{\beta',\gamma'=1}^{N_{\rm rank}}\; Z^{{\rm sc},ik}_{\beta \beta'}(E)\;\bar\tau^{{\rm sc}}_{k,\beta'\gamma'}(E)\;X^{{\rm sc},kj}_{\gamma'n}(E).
\label{eq:2.13b}
\end{eqnarray}
In this work we diagonalize Eqs.~(\ref{eq:2.13a}) and (\ref{eq:2.13b}) in a Coulomb basis to arrive at the transition amplitudes $X^{ij}_{mn}(q_i,q_j)$ needed to evaluate cross sections for the various three-body processes.
A detailed discussion on the evaluation of the effective potentials $Z^{{\rm sc},ij}_{mn}(E)$ and the diagonalization of the resulting
Faddeev-AGS equations in the Coulomb basis can be found in Ref.~\cite{hlophe:2022}.
\subsection{Microscopic calculation of nucleon-nucleus potentials in momentum space.}
\label{formal:ai}
The ingredients for the three-body Faddeev calculations discussed in Sec.~\ref{formal:3b} are the momentum space microscopic pairwise $N$-$A$ potentials (${\cal W}_{NA}$) in addition to the $n$-$p$ interaction. To compute ${\cal W}_{NA}$, we start from 
the microscopic ($A$+$1$)-body Hamiltonian, which in general includes two- and three-nucleon (NN and 3N) forces,
\begin{equation}
H = \sum\limits_{i<j=2}^{A+1}\frac{( \vec{k}_i-\vec{k}_j)^2}{2m_N}+\sum\limits_{i<j=2}^{A+1} V^{\rm NN}_{ij}+ \sum\limits_{i<j<k=3}^{A+1} V^{\rm 3N}_{ijk},
\label{eq:1.1}
\end{equation} 
 where ${\vec k}_i$ is the momentum of the $i^{\rm th}$ nucleon and $m_N$ is the nucleon mass. The many-body Hamiltonian can be recast as
 \begin{equation}
H = H^{(A)}+\frac{p^2}{2\mu}+{\bar V}^{\rm c}(r)+V^{\rm rel},
\label{eq:1.2}
\end{equation} 
 where the intrinsic Hamiltonian $H^{(A)}$ governs the dynamics of the $A$-nucleon target within the $N$+$A$ system, with ${p}$ and $\mu$ being the relative momentum and reduced mass of the $N$+$A$ system. 
Here ${\bar V}^{\rm c}(r)=\frac{1}{2}(1+\tau^{z}_{A+1})Z e^2/r$ is the average $N$-$A$ Coulomb potential (with $Z$ and $\tau^z_{A+1}$ representing, respectively, the atomic number of the target and twice the isospin projection of the nucleon), and the relative nucleon-nucleus interaction  is given by
\begin{equation}
V^{\rm rel} = \sum\limits_{i=1}^{A} V^{\rm NN}_{i,A+1}+ 
\sum\limits_{i<j=2}^{A}
V^{\rm 3N}_{ij,A+1}-{\bar V}^{\rm c}(r). 
\label{eq:1.3}
\end{equation} 

In momentum space, the NCSM/RGM ansatz~\citep{Quaglioni:2009mn} for the $N$+$A$ scattering wave function for a given spin-parity and isopsin (${\cal I}^\pi T$) takes the form  
\begin{equation}
\big|\Psi_{\nu_0;\;p_0}^{{\cal I}^\pi T}\big\rangle =  \sum\limits_{\nu}\;\int\;dp p^2\; \chi_{\nu\nu_0}^{{\cal I}^\pi T}(p,p_0)\;{\cal A}_{\nu}\;\big|\Phi_{\nu p}^{{\cal I}^\pi T}\big\rangle,
\label{eq:1.4}
\end{equation} 
where $\nu$ is a collective index denoting an arbitrary angular momentum channel while $\nu_0$ and $p_0$ indicate the incident channel and momentum. 
If we chose $I_A$, $\pi_A$, $T_A$, and $\alpha_A$ to denote respectively the spin, parity, isospin, and additional quantum numbers of the target nucleus, the channel states 
are given by
\begin{eqnarray}
\big|\Phi_{\nu p }^{{\cal I}^\pi T}\big\rangle=\left[\left(|A\; \alpha_A I^{\pi_A} T_A\big\rangle\Big|1\; \frac{1}{2}^{+}\;\frac{1}{2}\Big\rangle\right)^{sT}Y_{\ell}({\hat p}_{A,A+1})\right]^{{\cal I}^\pi T}\;\frac{\delta({p-p_{A,A+1}})}{pp_{A,A+1}},
\label{eq:1.5}
\end{eqnarray}
where $|\; \alpha_A I^{\pi_A} T_A\big\rangle$ is an eigenstate of the 
intrinsic Hamiltonian $H^{(A)}$ with energy eigenvalue 
$E_{\nu}^{(A)}$, $s$ the total spin, and $\ell$ the relative orbital angular momentum for the $N$+$A$ system in channel $\nu$. The eiegenstate is obtained by diagonalizing $H^{(A)}$ in the model space spanned by the NCSM $N_{\rm max}\hbar\Omega$ harmonic oscillator (HO) basis. Here $N_{\rm max}$ is the maximum number of HO quanta above the minimum energy configuration of the nucleons and $\Omega$ is the HO frequency. The operator ${\cal A}_\nu$
antisymmetrizes the incident nucleon with respect to the nucleons in the target $A$. The angular motion of the projectile nucleon with respect to the target nucleus is described by ${ Y}_{\ell} (\hat p_{A,A+1})$ while the relative momentum is given by
\begin{eqnarray}
\vec{p}_{A,A+1}= p_{A,A+1}\;\hat{p}_{A,A+1}=\frac{A}{A+1}\;\left[ \vec{k}_{A+1}-\frac{1}{A}\;\sum\limits_{i=1}^{A} \vec{k}_{i}\right].
\label{eq:1.6}
\end{eqnarray}

By introducing the momentum-space norm  ${\cal N}_{\nu\nu'}^{{\cal I}^\pi T}(p,p')\equiv \big\langle\Phi_{\nu p}^{{\cal I}^\pi T}\big|{\cal A_\nu}{\cal A_\nu'}\big|\Phi_{\nu' p' }^{{\cal I}^\pi T}\big\rangle$ and adopting the orthogonalization procedure of Ref.~\citep{Quaglioni:2009mn}, the ($A$+$1$)-body Schr\"odinger equation can be reduced into the binary form 
\begin{eqnarray}
\Bigg[E_{A+1}-E^{(A)}_{\nu}-\frac{p^2}{2\mu}\Bigg]\; \chi_{\nu\nu_0}^{{\cal I}^\pi T} (p,p_0)=\sum\limits_{\nu'}\int dp' p'^2\; {\cal W}_{\nu\nu'}^{{\cal I}^\pi T}(p,p')\; \chi_{\nu'\nu_0}^{{\cal I}^\pi T} (p',p_0),
\label{eq:1.7}
\end{eqnarray}
where $\chi_{\nu'\nu_0}^{{\cal I}^\pi T}(p',p_0)$ is the amplitude of relative motion,
${\cal W}_{\nu\nu'}^{{\cal I}^\pi T}(p,p')$ the effective two-body 
$N$-$A$
potential and $E_{A+1}$ the total energy of the $N$+$A$ system. 
The $N$-$A$ potential 
\begin{eqnarray}
{\cal W}_{\nu\nu'}(p,p')&=& \bar{\cal H}_{\nu\nu'}^{\rm mod}(p,p')+{\cal W}_{\nu\nu'}^{\rm eev}(p,p')+{\cal W}_{\nu\nu'}^{\rm kin}(p,p')+\bar{\cal V}^{\rm coul}_{\nu\nu'}(p,p'),
\label{eq:napot_full}
\end{eqnarray}
contains contributions from: $i)$ the orthogonalized Hamiltonian kernel inside the HO model space ($\bar{\cal H}_{\nu\nu'}^{\rm mod}$), $ii)$ the energy eigenvalue of the target nucleus (${\cal W}_{\nu\nu'}^{\rm eev}$), $iii)$ the relative kinetic energy (${\cal W}_{\nu\nu'}^{\rm kin}$), and $iv)$ the average Coulomb potential ($\bar{\cal V}^{\rm coul}_{\nu\nu'}$), 
where we have dropped the $J^\pi T$ superscript for brevity. Details of the derivation of the expressions for ${\cal W}_{\nu\nu'}^{{\cal I}^\pi T}(p,p')$ are presented in Appendix~\ref{appendixA}. 

The corresponding $t$~matrix that serves as input to the Faddeev-AGS equations and is needed for computing $N$+$A$ scattering phase shifts satisfies the momentum space Lippmann-Schwinger (LS) equation
\begin{eqnarray}
t_{\nu\nu_0}^{{\cal I}^\pi T}(p,p_0;E_{\rm kin})=  {\cal W}_{\nu\nu_0}^{{\cal I}^\pi T}(p,p_0)+\sum\limits_{\nu'} \; \int dp'\;p'^2\;{\cal W}_{\nu\nu'}^{{\cal I}^\pi T}(p,p')\;G_{0\nu}(E_{\rm kin},p')\;t_{\nu\nu_0}^{{\cal I}^\pi T}(p',p_0;E_{\rm kin}),
\label{eq:1.9}
\end{eqnarray}  
which is equivalent to Eq.~(\ref{eq:1.7}) complemented by the scattering boundary conditions, and the propagator is given by $G_{0\nu}(E_{\rm kin},p')=[E_{\rm kin}-\epsilon_{\nu}-{p'^2}/{2\mu}+i0]^{-1}$. Here $E_{\rm kin}\equiv E_{A+1}-E_{\nu_0}^{(A)}$ is the relative kinetic energy of $N+A$ system corresponding to the incident channel while $\epsilon_{\nu}\equiv E_{\nu}^{(A)}-E_{\nu_0}^{(A)}$ is the excitation energy of the target nucleus. As already noted in Sec.~\ref{formal:3b}, the Faddev calculations require a separable representation of the short-ranged component of the potential ${\cal W}_{\nu\nu'}^{{\cal I}^\pi T}(p,p')$. This is achieved by applying the
Ersnt-Shakin-Thaler~\cite{Ernst:1973zzb,Ernst:1974up} (EST) scheme which has been extensively verified for both nucleon-nucleus~\cite{Hlophe:2013xca,Hlophe:2014xca,Hlophe:2015rqn} and deuteron-nucleus systems~\cite{Hlophe:2017bkd,Hlophe:2019wwg}. We also note that, unlike typical phenomenological potentials based on local Woods-Saxon (WS)
functions, microscopic NCSM/RGM pairwise potentials employed in this work are instrinsically separable because they are given by expansions in the radial wave functions for $p$ and $p'$ separately. The EST scheme is merely employed to arrive at a specific separable form for ${\cal W}_{\nu\nu'}(p,p')$ that is convenient for solving the Faddeev-AGS equations and an accurate representation is attained using a small basis size.
\section{Results and Discussion}
\label{discuss}
\subsection{Momentum-space microscopic $n+^4{\rm He}$ and $p+^4{\rm He}$  potentials}
\label{results:nalpha}
\begin{figure}[ht]
\begin{center}
\includegraphics[width=15cm]{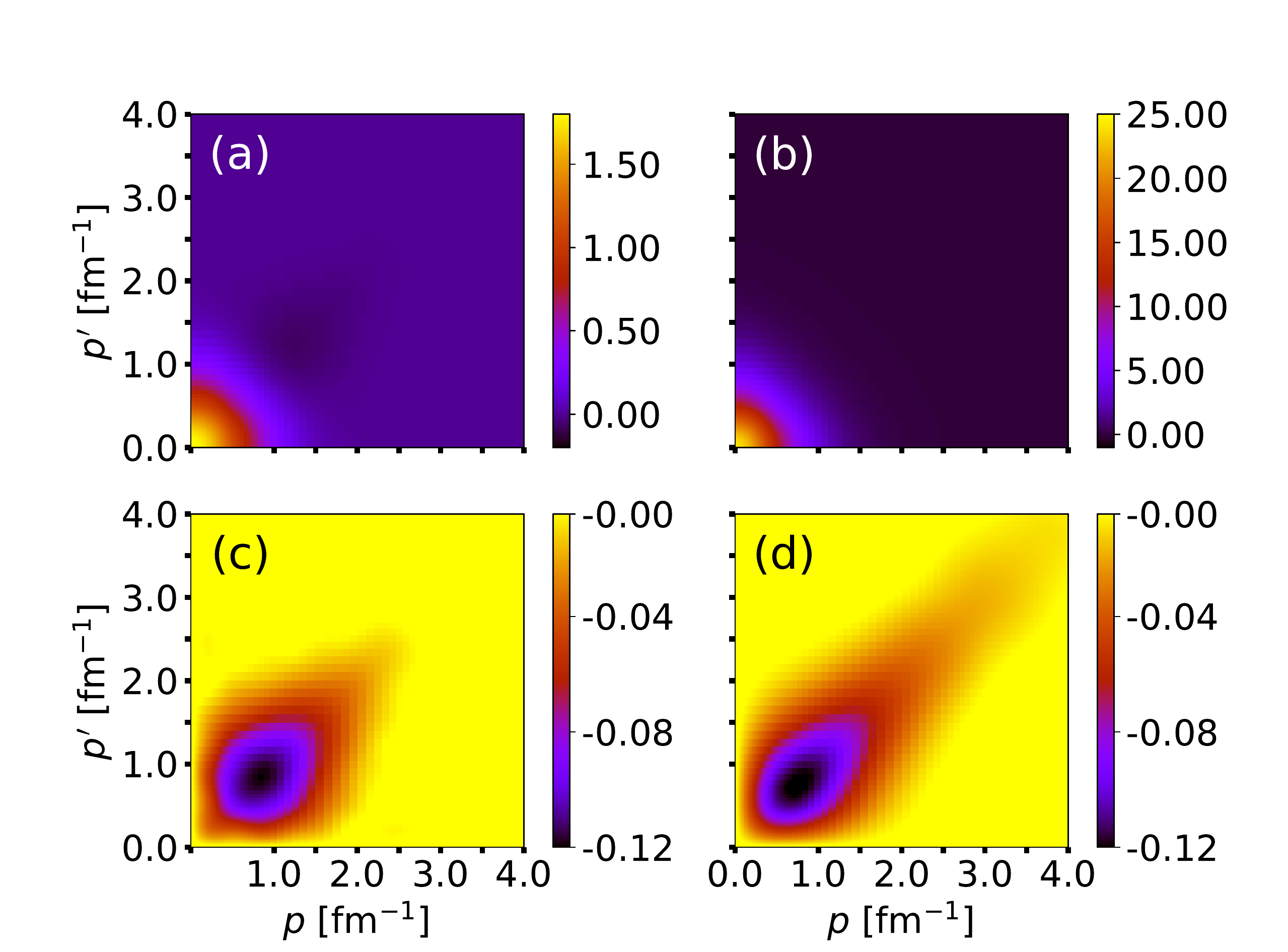}
\vspace{3mm}
\caption{[Color online] The partial wave $n$-$\alpha$ potential ${W_l^{\cal I}}(p,p')$ in units of fm$^2$ as a function of the momenta $p$ and $p'$. Panels (a) and (c) depict, respectively, the $s_{1/2}$ and $p_{3/2}$ microscopic NCSM/RGM potentials while the corresponding Woods-Saxon (WS) potential is shown in panels (b) and (d), respectively. The WS potential is scaled down by the strength of the Pauli projector~\cite{Hlophe:2017bkd} $\Gamma= 1100$~fm$^{-1}$. The NCSM/RGM model space is truncated at $N_{\rm max}=12/13$ for the positive/negative parity partial waves with $\hbar\Omega=14$~MeV and $\lambda_{\rm SRG}=1.5$~fm$^{-1}$.} 
\label{fig2}
\end{center}
\end{figure}
\begin{figure}[ht]
\begin{center}
\includegraphics[width=10cm]{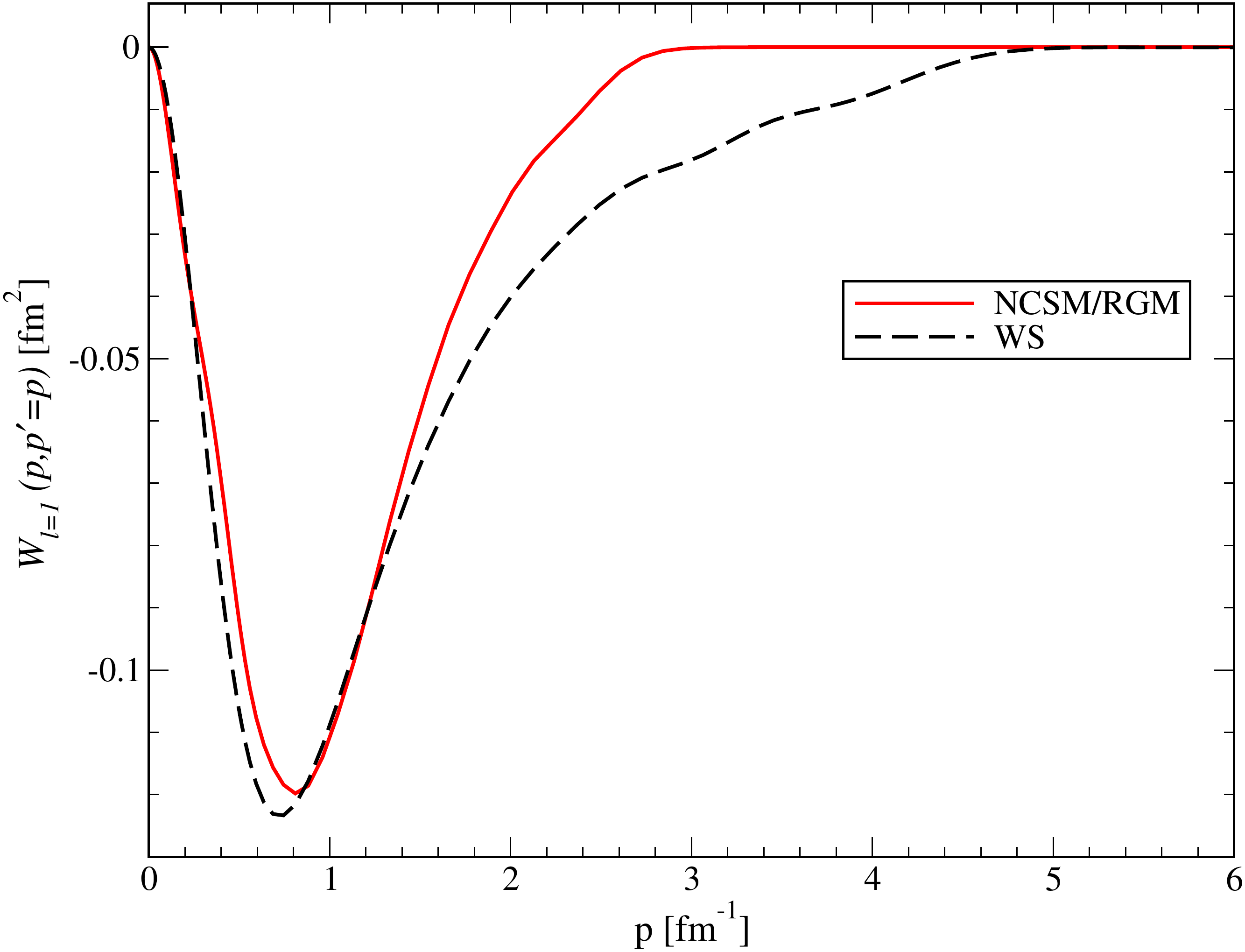}
\vspace{3mm}
\caption{[Color online] The diagonal elements of the $p$-wave potential ${W}_{\ell=1}^{{\cal I}}(p,p'=p)$ for the $n+\alpha{\rm }$ system as a function of the relative momentum coordinate 
$p$. The [red] solid lines depicts the microscopic potentials computed with the NCSM/RGM while the corresponding Woods-Saxon (WS) phenomenological potential~\cite{Bang:1983xpz} is indicated by [black] the dashed line. The NCSM/RGM model space is truncated at $N_{\rm max}=13$ with $\hbar\Omega=14$~MeV and $\lambda_{\rm SRG}=1.5$~fm$^{-1}$.}
\label{fig3}
\end{center}
\end{figure}
\begin{figure}[ht]
\begin{center}
\includegraphics[width=10cm]{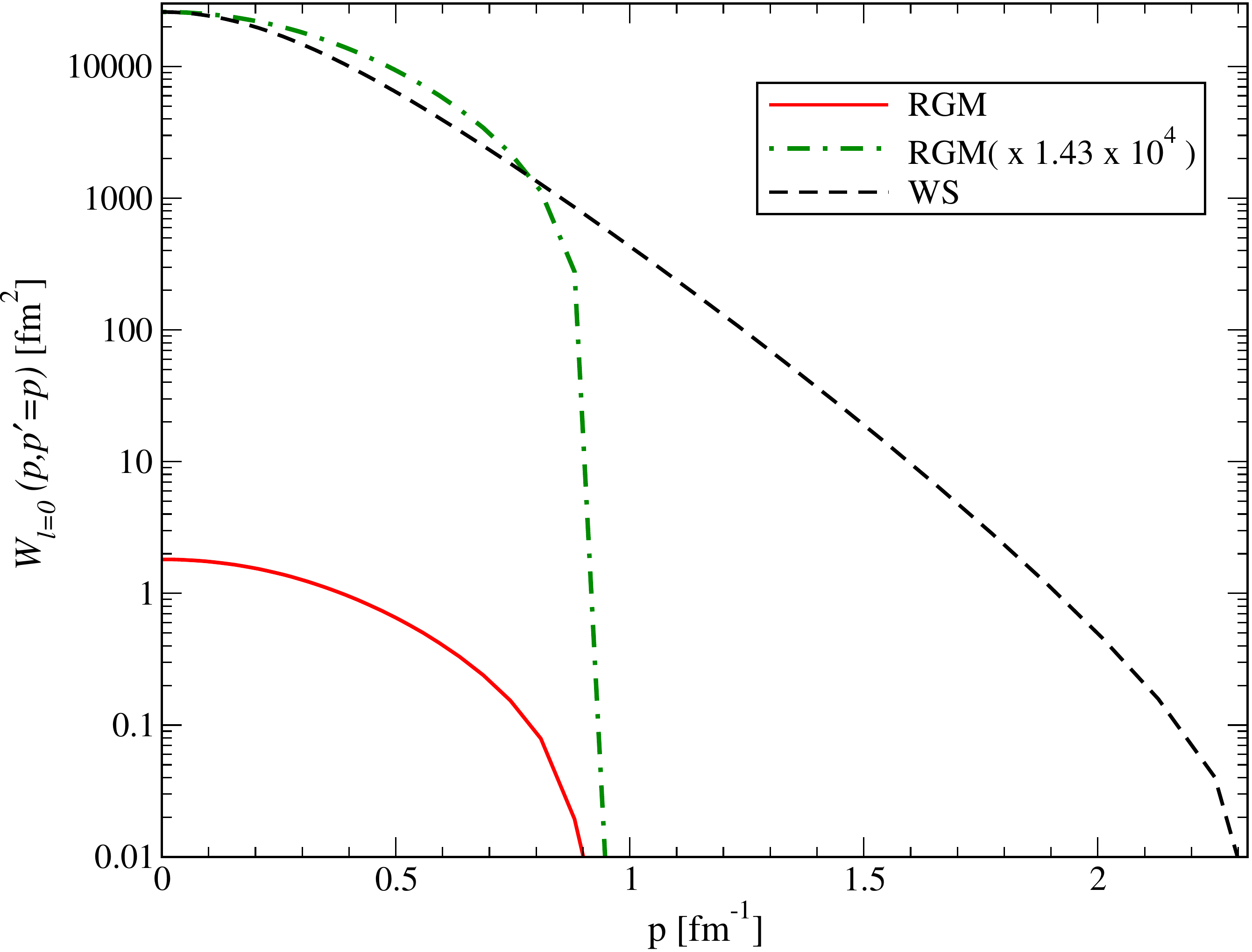}
\vspace{3mm}

\caption{[Color online] The diagonal elements of the $s_{1/2}$ potential ${W}_{\ell=0}^{{\cal I}}(p,p'=p)$ for the $n+\alpha{\rm }$ system as a function of the relative momentum coordinate 
$p$. The the microscopic potentials computed with the NCSM/RGM are depicted by [red] solid lines while the Woods-Saxon (WS) phenomenological potentials~\cite{Bang:1983xpz} are indicated by [black] dashed lines. A scaled up (by a factor of $1.43\times 10^{4}$) NSM/RGM potential is indicated by the [green] dash-dotted line. The NCSM/RGM model space is truncated at $N_{\rm max}=12$ with $\hbar\Omega=14$~MeV and $\lambda_{\rm SRG}=1.5$~fm$^{-1}$.}
\label{fig4}
\end{center}
\end{figure}
\begin{figure}[ht]
\begin{center}
\includegraphics[width=10cm]{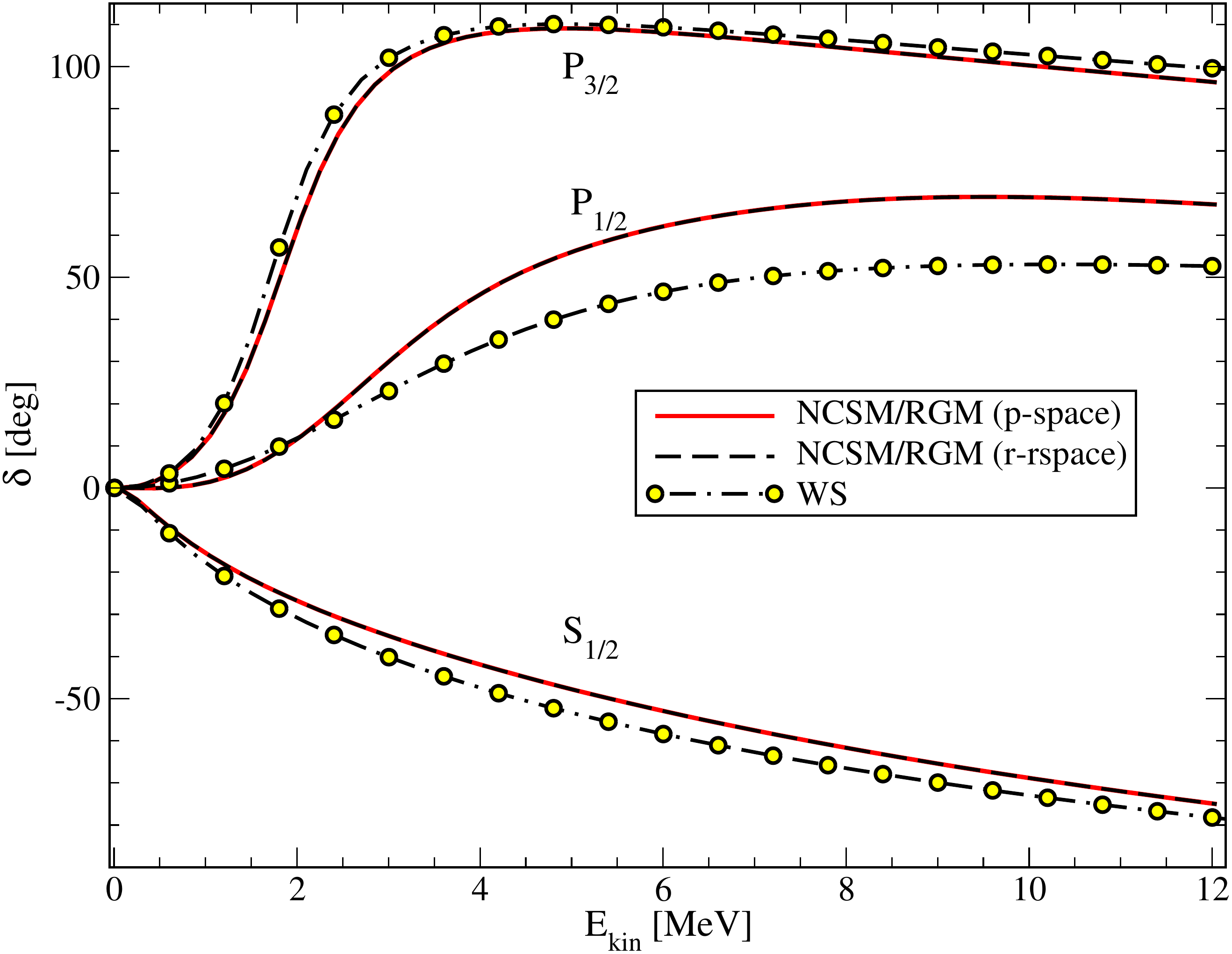}
\vspace{3mm}
\caption{[Color online] The phase shifts for $n+\alpha{\rm }$ scattering as a function of the
relative kinetic energy. The [red] solid lines show the phase shifts computed in momentum space using the LS
equation while the [black] dashed lines represent the corresponding coordinate space results obtained via the
microscopic R-matrix formalism~\cite{Hesse:1998nha,Hesse:2002vvv}. The dash-dotted lines with circles indicate
the phase shifts computed using a phenomenologically fitted Woods-Saxon (WS) potential~\cite{Bang:1983xpz}. The
NCSM/RGM model space is truncated at $N_{\rm max}=12/13$ for the positive/negative parity partial waves. Note that
the coordinate space and momentum space NCSM/RGM results are indistinguishable within the line thickness.}
\label{fig5}
\end{center}
\end{figure}
The starting point for all our calculations is a
soft SRG-evolved NN potential at fourth order (SRG-N$^3$LO NN potential) obtained by evolving the $\chi$EFT
interaction of Ref.~\cite{Entem:2003ft} to a momentum resolution scale of $\lambda_{\rm SRG}=1.5$ fm$^{-1}$. The low momentum resolution scale helps to speed up convergence of the HO expansion and ensures a reasonable description of $^6$Li bound state properties despite the omission of explicit 3N forces. For all calculations presented in this section the wave function of the $\alpha$-particle is evaluated using an HO basis expansion up to $N_{\rm max}=12$ while the positive/negative-parity NCSM/RGM
partial-wave potentials are computed in momentum space using $N_{\rm max}=12/13$ for the $N$-$\alpha$ system. This model space, together with the HO frequency $\hbar\Omega=14$~MeV, was demonstrated to yield an accurate convergence for the
$d$-$\alpha$ phase shifts for an identical choice of the NN interaction~\cite{Navratil:2011ay}.

To obtain the microscopic $N$-$\alpha$ momentum-space potentials
required as input for the Faddeev $n$+$p$+$\alpha$ calculation, we apply the momentum-space NCSM/RGM approach described in Section~\ref{formal:ai}. For this case, the $N$-$\alpha$ potentials are
diagonal in $\ell$ and the isospin is fixed at $T=1/2$, or ${\cal W}_{\nu\nu'}^{{\cal I}^\pi T}\equiv W_{\ell}^{{\cal I}}\;\delta_{\nu\nu'}$, since the ground state spin and isospin of the $\alpha$ particle are zero. 
First, we note that the NCSM/RGM-generated $N$-$\alpha$ potentials are smooth short-ranged functions of the momentum coordinates (Fig.~\ref{fig2}) and are therefore suitable for implementation in the Faddeev-AGS framework to compute $d$+$\alpha$ observables. Further, we see that the microscopically computed $p$-wave potential (Fig.~\ref{fig3}) has a similar shape to the phenomenologically fitted Woods-Saxon (WS) potential~\cite{Bang:1983xpz} at low momenta. However, there is a bigger difference at large momenta owing to the presence of high-momentum components in the WS. A similar comparison for the $s$-wave (Fig.~\ref{fig4}) shows that the shape of the microscopically-computed potentials is similar to its WS counterpart. However, similarly to the $p$-wave potential, the $s$-wave potential NCSM/RGM potential lacks the high momentum components owing to the low-momentum resolution of the NN potential used in the microscopic calculation. We also note that the magnitude of the WS $s$-wave potential is orders of magnitude larger than the NCSM/RGM potential. This is a consequence of the approximate treatment of the Pauli exclusion principle in the WS potential~\cite{Hlophe:2017bkd}. Specifically, the WS well used to fit the $N$+$\alpha$ phase shifts supports a Pauli-forbidden bound state that is removed by adding the term $|\phi_0\rangle\Gamma\langle \phi_0|$, with $|\phi_0\rangle$ being the bound state wave function and the parameter $\Gamma\rightarrow \infty$ is the strength of the Pauli projector (in practice we work with finite values, e.g., $\Gamma=1100$~fm$^{-1}$ in this case). This procedure approximates the Pauli exclusion principle by effectively shifting the spurious bound state to very large positive energies, well beyond the kinematic range under consideration, leading to the unusally large values for the $s$-wave potential.

To verify the accuracy of the NCSM/RGM-generated momentum space potentials, we employ the LS
equation~\eqref{eq:1.9} to compute  $N$-$\alpha{\rm }$ scattering phase shifts and compare them with
the results obtained by solving the orthogonalized NCSM/RGM equations in coordinate space, using the
microscopic R-matrix method on a Lagrange mesh~\cite{Hesse:1998nha,Hesse:2002vvv}. The agreement between the
two calculations is excellent (Fig.~\ref{fig5}), and demonstrates the accuracy of the momentum space
$N$-$\alpha$ potentials. We also compare the results of the microscopic potentials to the phase shifts
computed using the phenomenological WS fits. There is generally a reasonable agreement for the $s_{1/2}$ and
$p_{3/2}$ partial waves. The $s_{1/2}$ phase shifts are substantially larger than those of the WS potential,
indicating an underestimation of the spin-orbit splitting, as already observed in Ref.~\cite{Navratil:2011ay}. The inclusion of three-nucleon forces combined with the solution of the problem within the more complete approach of the no-core shell model with continuum (NSCMC) resolves this issue~\cite{Navratil:2016ycn,Hupin:2014kha}.             

\subsection{NCSM/RGM calculations for $^6$Li and $d$-$^4$He scattering.}
\label{ncsm/rgm}

We obtain the fully microscopic solution for the $d$-$\alpha$ bound and scattering states using the
NCSM/RGM approach for deuteron-nucleus systems~\cite{Navratil:2011ay}. To enable a direct comparison with the
Faddeev $n$+$p$+$\alpha$ calculations, we employ the same NN interaction and consistent HO model space as for
the $N$-$\alpha$ potentials (Sec.~\ref{results:nalpha}). Different from the Faddeev formalism, here the wave
fuction of the deuteron projectile is computed by diagonalizing the two-nucleon intrinsic Hamiltonian in a
two-body NCSM model space of consistent size $N_{\rm max}$ and HO frequency $\Omega$ as the $\alpha$ particle.
Further, the deformation and virtual breakup of the weakly-bound deuteron projectile are treated by including
excited deuteron pseudo-states (discretizing the deuteron continuum) in the NCSM/RGM coupled-channel equations,
similar to what was done in Ref.~\cite{Navratil:2011ay}. In that work it was demonstrated that the inclusion of
the $n+p$ angular momentum channels $^3S_1-^3D_1$, $^3D_2$, and $^3D_3-^3D_3$, while limiting the number of
deuteron pseudo-states to no more than seven, yields well-converged results for the $d+\alpha$ system.

\subsection{Faddeev calculations for the $d+^4{\rm He}$ system}
\label{results:faddeev}
We proceed to solve the bound state Faddeev equations for the $n$+$p$+$\alpha$ system using the effective three-body Hamiltonian of Eq.~(\ref{eq:2.1}) where ${\cal W}_{n\alpha}$, and ${\cal W}_{p\alpha}$ are computed
microscopically as described in Sec.~\ref{results:nalpha}. By varying the value of $N_{\rm max}$ used in the
computation of the $N-\alpha$ potentials, we determine that the $^6$Li ground state energy is indeed
well-converged at $N_{\rm max}=12$ (Fig.~\ref{fig4}), which is consistent with the convergence patterns of the
$d$+$\alpha$ microscopic calculations discussed in Sec.~\ref{ncsm/rgm}. Additionally, the convergence of
the $^6$Li ground state energy with $N_{\rm max}$ is similar to that of the relative binding energies
$E_2\equiv E(^6{\rm Li})-E(\alpha)-E(d)$ and $E_3\equiv E(^6{\rm Li})-E(\alpha)$. Further, the Faddeev calculation requires a truncation in the number of 
$N$-$\alpha$ partial waves included in the effective three-body Hamiltonian. We find that the $^6$Li ground state is well-converged for the total angular momenta ${\cal I}_{n{\rm +}p}\le3$ and ${\cal I}_{N{\rm -}\alpha}\le5/2$ for
the $n$+$p$ and $N$+$\alpha$ subsystems, respectively. The  $n$+$p$+$\alpha$ three-body binding energy is determined to be $E_3 = -3.18$~MeV which is combined with the energy of the $\alpha$-particle (Table~\ref{table1}) to obtain a $^6$Li ground state energy of $E_{6}=-31.41$~MeV. The resulting $^6$Li ground
state energy is found to be $E_{6}=-32.01$~MeV which is approximately $600$~KeV more bound than the Faddeev three-body calculation. This underbinding of the Faddeev calculation indicates a missing attractive strength in
the $n-p-\alpha$ interaction that is not captured by the sum of the pairwise $n-\alpha$ and
$p-\alpha$ potentials. Since the $\alpha$ particle is fixed in its ground state in both the NCSM/RGM and
Faddeev calculations, this effect can be attributed to the antisymmetrization of the six nucleon problem.
Indeed, a close inspection of the NCSM/RGM deuteron-nucleus potential (see e.g., Eqs.~(B2a)$-$(B2f) of
Ref.~\cite{Navratil:2011ay}) reveals the presence of irreducible three-body terms that arise from a
simultaneous exchange of the two nucleons inside the deuteron with nucleons in the $\alpha$ particle.      
\begin{figure}[ht]
\begin{center}
\includegraphics[width=10cm]{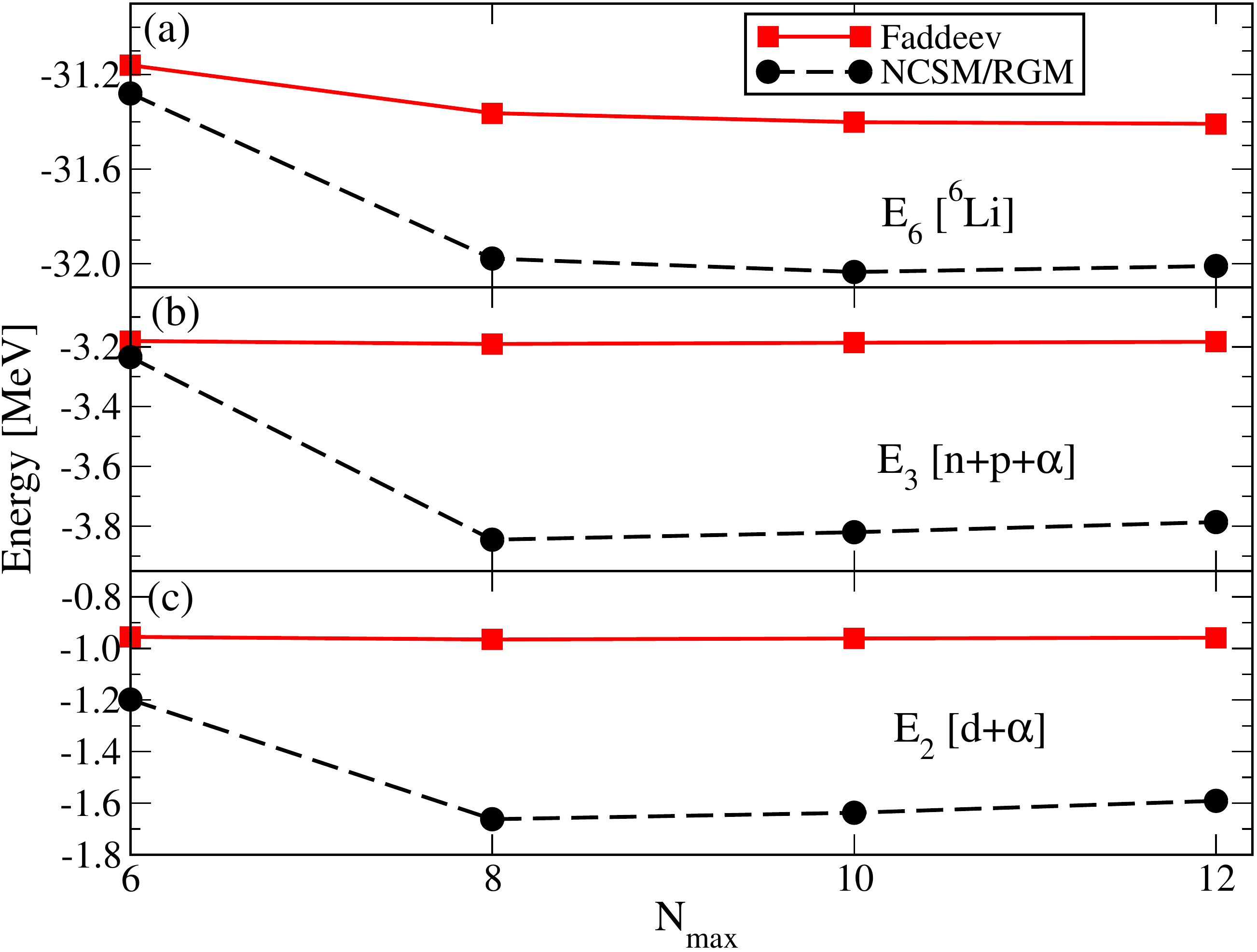}
\vspace{3mm}
\caption{[Color online] The $^6$Li ground-state binding energy as a function of $N_{\rm max}$ calculated using
the NCSM/RGM and the three-body Fadeev approach. Panel (a) depicts the total 6-body binding energy of $^6$Li
while panels (b) and (c) respectively show the relative binding energy for the $n+p+\alpha$ and $d+\alpha$
systems. The Faddeev results are indicated by [red] solid lines with filled squares while the NCSM/RGM calculations
are depicted by [black] dashed lines with circles. The NCSM/RGM model space is truncated at $N_{\rm max}=12/13$ for the
positive/negative parity partial waves and $\hbar\Omega=14$~MeV. The microscopic $N$-$\alpha$ potentials that was used as input in the Faddeev computation
of the $^6$Li bound state were obtained using $N_{\rm max}=12/13$ with the angular momentum truncation ${\cal I}_{n{\rm +}p}\le 2$ and ${\cal I}_{N{\rm +}\alpha}\le 5/2$.
}
\label{fig6}
\end{center}
\end{figure}
\begin{table}[ht]
\centering
\begin{tabular}{ccccccccccccc}
\hline\hline
       &&$\alpha$ (NCSM)&& $d$ (NCSM)&& $d$ (LS)&& $^6$Li (NCSM/RGM) && $^6$Li (Faddeev)\\
 \hline
E[MeV] &&-28.22 && -2.20&& -2.22&& -32.01 &&-31.41\\
\hline\hline
\end{tabular}
\caption{The ground state energies for the $d$, $\alpha$, and $^6$Li computed using the SRG-N$^3$LO
potential~\cite{Entem:2003ft} with $\lambda_{\rm SRG}=1.5$ fm$^{-1}$. The NCSM results were obtained using
$N_{\rm max}=12$. The binding energy of the deuteron calculated by solving the LS equation in momentum space
for the $np$ system is also listed. The angular momentum channels $^3S_1-^3D_1$, $^3D_2$, and $^3D_3-^3D_3$ of
the $np$ system were included in the NCSM/RGM calculation while limiting the number of deuteron pseudostates to
no more than seven. The microscopic $N$-$\alpha$ potentials that was used as input in the Faddeev computation
of the $^6$Li bound state were obtained using $N_{\rm max}=12/13$ with the angular momentum truncation ${\cal I}_{n{\rm +}p}\le 2$ and ${\cal I}_{N{\rm +}\alpha}\le 5/2$.           }
\label{table1}
\end{table}

Next, we solve the Faddeev-AGS equations for $d+\alpha$ scattering at energies below the deuteron breakup threshold, starting from the same three-body Hamiltonian as in the bound state calculations above. We first
illustrate the convergence of the Faddeev calculations with respect to the number $n$+$p$ and $N$+$\alpha$
partial waves, and determine that a maximum angular momentum of ${\cal I}_{n{\rm +}p}=3$ and ${\cal
I}_{N-\alpha}=9/2$ is sufficient to reach convergence, respectively (Fig.~\ref{fig6}). At the same time, we
employ the NCSM/RGM to microscopically compute $d+\alpha$(g.s) elastic scattering and proceed to compare the
results with the Faddeev calculations for the two dominant $d+\alpha$(g.s) partial waves at these energies. We
find that there is a relatively small difference between the $^3S_1$ phase shifts computed using the Faddeev and NCSM/RGM methods. However, there is a significant difference in the $^3D_3$ phase shifts which manifests
primarily as a shift in the position of the $3^+$ resonance. The Faddeev calculations yield a resonance energy
that is approximately $400$~keV larger relative to the NCSM/RGM result, which is consistent with the
underbinding observed in the aforementioned ground state calculations of $^6$Li. We note that the $^3D_2$ resonance is also shifted higher in energy and lies beyond the kinematic range considered in this work. Lastly,
we compute energy and angular distributions in order to quantify the effects of the irreducible three-body force at the level of scattering observables. The energy distributions are dominated by the $^3D_3$ resonance and,
as expected, the Faddeev calculation exhibits a peak that is approximately $400$~keV higher than that of the NCSM/RGM (Fig.~\ref{fig8}). Also, the differences between the angular distributions are largest for energies
closest to the resonance. The effect of the omitted 3BF arising due to Pauli effects is thus sizeable for both
the bound state and scattering observables of the $d+\alpha$ system. Moreover, this missing 3BF is attractive
in the $d+\alpha$ partial waves considered here. Since the shapes of angular distributions, e.g. in transfer reactions, are used to extract spectroscopic information, the quality of such information can be impacted by the 3BF effects.                
\begin{figure}[ht]
\begin{center}
\includegraphics[width=10cm]{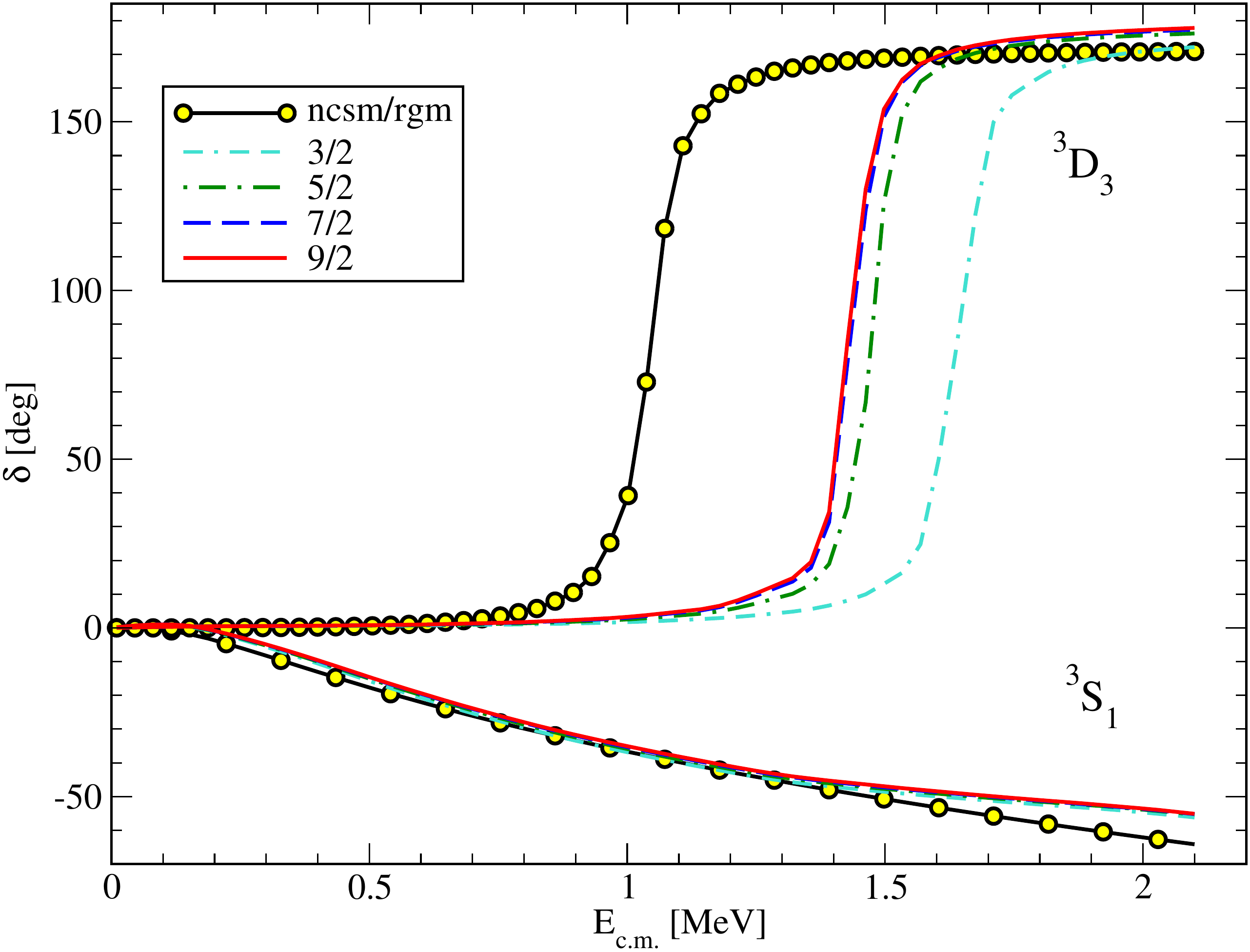}
\vspace{3mm}
\caption{[Color online] The phase shifts for elastic $d+\alpha$ scattering as a function of the center-of-mass energy for different partial wave three-body model spaces. The [black] solid line with filled circles shows phase
shifts computed using the NCSM/RGM. The Faddeev results are depicted by [red] solid, [blue] dashed, [green] dash-dotted, and [turquoise] dash-dash-dotted lines for the partial wave model spaces ${\cal I}_{N{\rm +}\alpha}\le 9/2$,\;${\cal I}_{N{\rm +}\alpha}\le 7/2$,\;${\cal I}_{N{\rm +}\alpha}\le 5/2$,\; and ${\cal I}_{N{\rm +}\alpha}\le 3/2$ as indicated in the figure. The $n$+$p$ partial waves are limitted to ${\cal I}_{n{\rm +}p}\le 3$. 
The NCSM/RGM model space is truncated at $N_{\rm max}=12/13$ for the positive/negative parity partial waves and $\hbar\Omega=14$~MeV.   
}
\label{fig7}
\end{center}
\end{figure}
\begin{figure}[ht]
\begin{center}
\includegraphics[width=10cm]{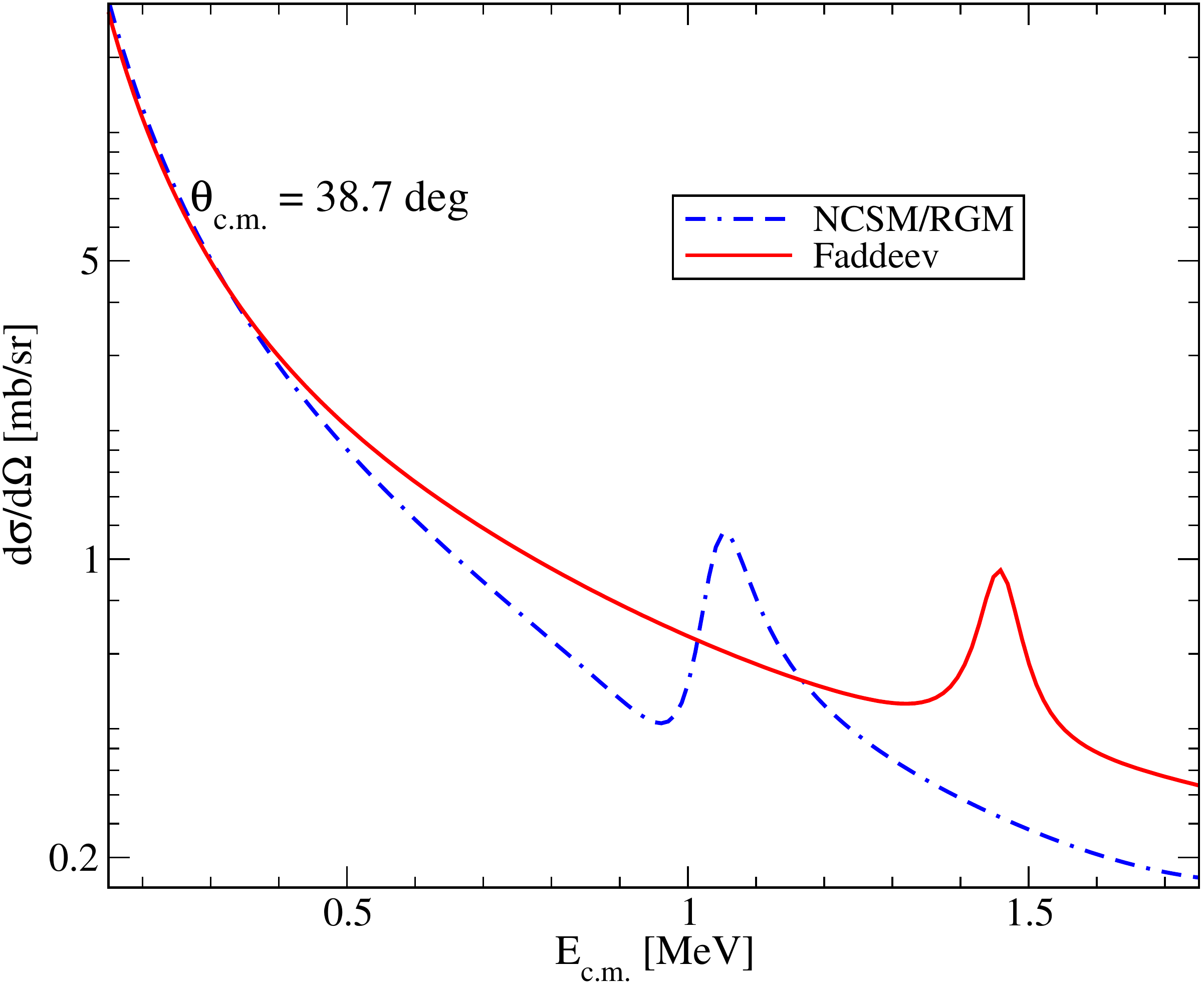}
\vspace{3mm}
\caption{[Color online] The differential cross section  for elastic $d+\alpha$ scattering as a function of the center-of-mass energy $E_{\rm c.m.}$ at the scattering angle $\theta_{\rm c.m.}=38.7$~deg. The [blue] dashed lines shows phase shifts computed using the NCSM/RGM while the Faddeev results are depicted by [red] solid lines. The
model space for the Faddeev calculation is restricted to a total two-body angular momentum of ${\cal I}_{n+p}\le3$ and ${\cal I}_{N{\rm +}\alpha}\le9/2$ for the $n$+$p$ and $N$-$\alpha$ subsystems. The NCSM/RGM model
space is truncated at $N_{\rm max}=12/13$ for the positive/negative parity partial waves and $\hbar\Omega=14$~MeV. 
}
\label{fig8}
\end{center}
\end{figure}
\begin{figure}[ht]
\begin{center}
\includegraphics[width=10cm]{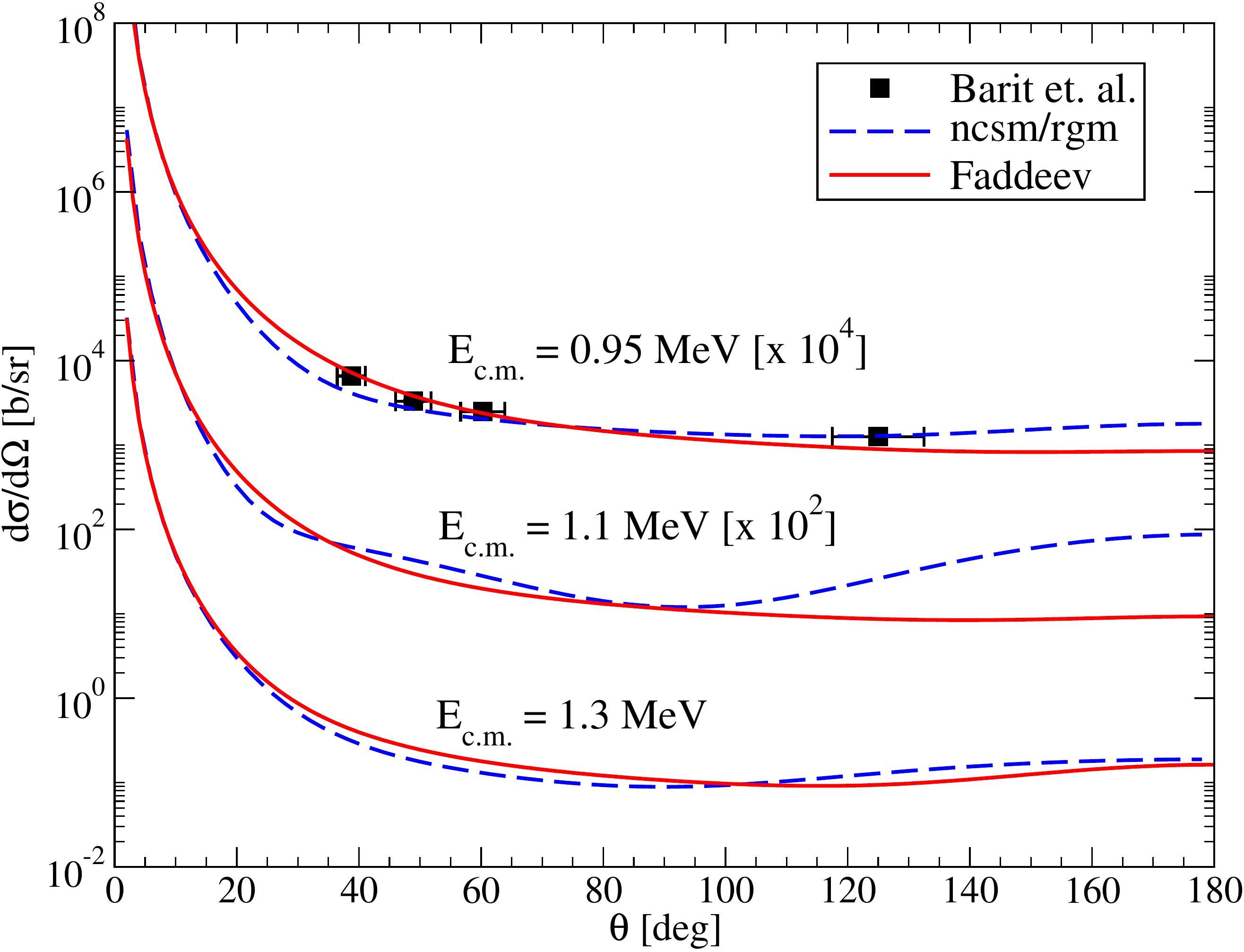}
\vspace{3mm}
\caption{[Color online] The differential cross section  for elastic $d+\alpha$ scattering as a function of the
center-of-mass angle $\theta_{c.m.}$ at varying energies $E_{c.m.}$. The solid lines shows phase shifts computed using the Faddeev method while the NCSM/RGM results are depicted by dashed lines. The model space for
the Faddeev calculation is restricted to a total two-body angular momentum of ${\cal I}_{np}\le3$ and ${\cal I}_{N-\alpha}\le9/2$ for the $np$ and $N$-$\alpha$ subsystems. Experimental data from~\cite{Barit:1979} are
included. The NCSM/RGM model space is truncated at $N_{\rm max}=12/13$ for the positive/negative parity partial waves and $\hbar\Omega=14$~MeV.  
}
\label{fig9}
\end{center}
\end{figure}
\section{Conclusion and Outlook}
\label{conclusion}
While the three-body model provides an indispensable tool for the description of deuteron-induced nuclear reactions, its predictive power is still limited by the lack of constraints on the effective three-body Hamiltonian. Particularly, the reduction of the many-body deuteron-nucleus problem into the three-body space results in an effective Hamiltonian that consists of an irreducible neutron-proton-nucleus 3BF in addition to the pairwise
nucleon-nucleus potentials. In this work we have studied the irreducible $n$-$p$-$\alpha$ 3BF arising from antisymmetrization effects and quantified its impact on observables of the $d$+$\alpha$ system. We started from a microscopic five-body Hamiltonian constructed using SRG-evolved N$^3$LO chiral NN potentials
and employed the NCSM/RGM~\cite{Quaglioni:2009mn} to compute momentum space  $N$+$\alpha$ potentials. To benchmark the potentials, we used the LS equation to obtain the corresponding scattering solutions and demonstrated that the resulting phase shifts are in excellent agreement with those obtained via the standard two-body NCSM/RGM by solving R-matrix equations in coordinate space. 
Next, we constructed an effective three-body potential for the $n+p+\alpha$ system by summing the pairwise potentials and proceeded to solve the corresponding Faddeev-AGS equations for the $^6$Li ground state as well as $d$-$\alpha$ scattering. At the same time, we carried out
corresponding microscopic calculations using the NCSM/RGM starting from a six-body microscopic Hamiltonian based on the same NN interaction as in the $N$+$\alpha$ case. 

We observed that the Faddeev calculation yields
a $^6$Li ground state that is approximately $600$~keV shallower than the one obtained  with the NCSM/RGM,
indicating an omitted attractive 3BF. Additionally, the $d+\alpha$ three-body calculations yield a $3^+$
resonance energy that is approximately $400$~keV larger compared to  the NCSM/RGM result, which is consistent with the observed underbinding of the $^6$Li ground state. The two methods also yield angular distributions
with different shapes due to differences in the position of the $3^+$ resonance. Since the $\alpha$ particle is
fixed in its ground state, the origin of this 3BF can be ascribed to two-nucleon exchange processes that are intrinsically three-body in nature.

In the future, three-body calculations of $d+\alpha$ may be improved by using the two-nucleon exchange terms in NCSM/RGM-generated $d-\alpha$ potential to evaluate the irreducible three-body force. Moreover, a
similar study based on  more complete formalism of the NCSMC~\cite{Hupin:2014iqa,Navratil:2016ycn} (where the NCSM/RGM ansatz is augmented with an expansion in square-integrable NCSM eigenstates of the composite ($A+2$)-nucleon system), is necessary for the quantification of additional
components arising from excitations of the nucleons inside the target nucleus. Lastly, performing similar studies for other light nuclei and establishing a mass dependence of the irreducible three-body force
would providevaluable information for phenomenological models that have a wider range of applicability in mass and energy.

\newpage
\begin{acknowledgments}
The authors would like to thank Chl\"oe Hebborn for useful discussions. Computing support for this work came from the Lawrence Livermore National Laboratory (LLNL) Institutional Computing
Grand Challenge program. This work was performed under the auspices of the U.S. Department of Energy by LLNL under contract number DE-AC52-07NA27344. This material is based upon work supported by the U.S. Department of Energy, Office of Science, Office of Nuclear
Physics, under Work Proposal No. SCW0498 and LLNL LDRD project No. 22-LW-003.
\end{acknowledgments}

\appendix
\section{Ab initio nucleon-nucleus potentials in momentum space }
\label{appendixA}
To compute the potential ${\cal W}_{\nu\nu'}^{{\cal I}^\pi T}(p',p)$, we first note that the norm kernel
consists of the identity operator and a short-ranged exchange term ${\cal N}_{\nu\nu'}^{{\rm ex, }\;{\cal I}^\pi T} (p',p)$,  so that ${\cal N}_{\nu\nu'}^{{\cal I}^\pi T} (p,p')=\delta_{\nu\nu'}\frac{\delta(p-p')}{pp'}+{\cal N}_{\nu\nu'}^{{\rm ex, }\; {\cal I}^\pi T} (p',p)$.
The superscript ${\cal I}^\pi T$ will be dropped from all terms hereafter for brevity. We proceed by defining the harmonic oscillator (HO) model space
\begin{eqnarray}
P_{\ell}\equiv \sum\limits _{n=0}^{n_{\rm max}}|R_{n\ell}\rangle\langle |R_{n\ell}|,
\label{eq:hospace}
\end{eqnarray}
and the complimentary space $Q_{\ell}\equiv   1-P_{\ell}$, where the basis
vectors are the radial harmonic oscillator wavefunctions $|R_{n\ell}\rangle$, with $n$ being the radial quantum
number. The second term of the norm kernel can be accurately represented within the model space $P$, owing to
its short-ranged character. Contrarily, the delta function has components within the HO model space as well as
in the complimentary space. By defining the component of the norm kernel inside the model space
$\Lambda_{n\nu,n'\nu'}$, we can express the momentum space norm kernel in the form
\begin{eqnarray}
{\cal N}_{\nu\nu'} (p,p')&=& \delta_{\nu\nu'}\;Q_{\ell}(p,p')+ [P\;\Lambda\;P]_{\nu\nu'}(p',p),\cr\cr &\equiv& \left[\frac{\delta(p-p')}{pp'}-\sum\limits_{n} \tilde{R}_{n\ell}(p)\tilde{R}_{n\ell}(p') \right]\delta_{\nu\nu'}+\sum\limits_{nn'} \tilde{R}_{n\ell}(p)\; \Lambda_{n\nu,n'\nu'}\; \tilde{R}_{n\ell'}(p'),
\label{eq:a1.1}
\end{eqnarray}
where the $\tilde{R}_{n\ell}(p)$ is the momentum space representation of the radial harmonic oscillator
wavefunction. Additionally,  we introduce the square root of the norm kernel ${\cal N}^{1/2}$ and its inverse
${\cal N}^{-1/2}$ which are defined such that $[{\cal N}^{1/2}\cdot {\cal N}^{1/2}]_{\nu\nu'} (p,p')\equiv
{\cal N}_{\nu\nu'} (p,p')$ and $[{\cal N}^{1/2}\cdot{\cal N}^{-1/2}]_{\nu\nu'} (p,p') \equiv
\frac{\delta(p,p')}{pp'}\delta_{\nu\nu'}$. The multiplication by the norm implies a sum over channel indices and an integral over the momentum coordinate such that, e.g., 
\begin{eqnarray}
[{\cal N}^{1/2}\cdot {\cal N}^{1/2}]_{\nu\nu'} (p,p')=\sum\limits_{\nu''}\int\;dp'' p''^2\;{\cal N}_{\nu\nu''}^{\rm 1/2}(p,p'')\;{\cal N}_{\nu''\nu'}^{1/2}(p'',p').
\label{eq:a1.2}
\end{eqnarray}
Further, the (inverse) square root fulfills Eq.~(\ref{eq:a1.1}), with matrix elements $\Lambda_{n\nu,n'\nu'}^{\pm1/2}$ replacing $\Lambda_{n\nu,n'\nu'}$ and represents the component of the norm square (inverse square) root inside the $P$ space.         
To proceed, we note that the momentum space representation of the average Coulomb potential $\bar V^{\rm c}(r)$ is non-local and has an angular dependence so that ${\bar V}^{\rm c} \equiv {\bar V}^{\rm c}_{\ell}(p,p')$. By introducing the average Coulomb potential kernel ${\cal V}^{\rm c}_{\nu\nu'}(p,p')={\bar V}^{\rm c}_{\ell}(p,p')\;\delta_{\nu\nu'}$ and the relative potential kernel
\begin{eqnarray}
{\cal V}_{\nu\nu'}^{\rm rel}(p,p') \equiv \big\langle\Phi_{\nu p}\big|{\cal A}\;V^{\rm rel}{\cal A}\big|\Phi_{\nu' p'}\big\rangle,
\label{eq:a1.3}
\end{eqnarray}
the Hamiltonian kernel can be expressed as
\begin{eqnarray}
{\cal H}_{\nu\nu'}(p,p') = E_{\nu}^{(A)}\;{\cal N}_{\nu\nu'}(p,p')+\frac{p^2}{2\mu}\;{\cal N}_{\nu\nu'}(p,p')+\big[{\cal V}^{\rm c}\;{\cal N}\big]_{\nu\nu'}(p,p')+{\cal V}_{\nu\nu'}^{\rm rel}(p,p').
\label{eq:a1.4}
\end{eqnarray}
The $(A+1)$-body Schr\"odinger equation satisfied by the NCSM/RGM wave function of Eq.~(\ref{eq:1.3}) can thus be cast as
\begin{eqnarray}
\sum\limits_{\nu'}\int\;dp' p'^2\;\Big[E_{A+1}\;{\cal N}_{\nu\nu'}(p,p')-{\cal H}_{\nu\nu'}(p,p')\Big] \left[{\cal N}^{-1/2}\;\chi\right]_{\nu'\nu_0} (p',p_0)=0,
\label{eq:a1.5}
\end{eqnarray}
or in the condensed form
\begin{eqnarray}
\left[E_{A+1} -\bar{\cal H}\right]\;\chi =0,
\label{eq:a1.6}
\end{eqnarray} 
after multiplying by ${\cal N}^{-1/2}$ from the left, where the effective potential is given by $\bar{\cal H}\equiv{\cal N}^{-1/2}\;{\cal H}\;{\cal N}^{-1/2}$ and $E_{A+1}$ is the total energy of the 
nucleon-target system. The model space Hamiltonian kernel $ \bar{\cal H}^{\rm mod}$ (i.e., the component 
of the Hamiltonian kernel $\bar{\cal H}_{\nu\nu'}$ within the HO model space is computed similarly 
to the NCSM/RGM presented in Ref.~\cite{Quaglioni:2009mn} except that the momentum space HO wavefunctions are used in the place of their coordinate space counterparts. 
Further, there are corrections to $\bar{\cal H}_{\nu\nu'}(p,p')$ emerging from outside the model space $P$ due to the presence of long-ranged terms in Eq.~(\ref{eq:a1.4}), namely, the target nucleus energy eigenvalue ${\cal H}_{\nu\nu'}^{\rm eev}(p,p')\equiv E_{\nu}^{(A)}\;{\cal N}_{\nu\nu'}(p,p')$, (2) the relative nucleon-nucleus kinetic energy ${\cal H}_{\nu\nu'}^{\rm kin}(p,p')\equiv E_{\nu}^{(A)}\;{\cal N}_{\nu\nu'}(p,p')$, and
(3) the average Coulomb potential, ${\cal V}_{\nu\nu'}^{\rm coul}(p,p')\equiv \delta_{\nu\nu'}\;\left[\bar V_{\ell}^{\rm c}\;Q_{\ell}\right]_{\nu\nu'}(p,p')$. To compute the out-of-model-space contributions to $\bar{\cal H}_{\nu\nu'}(p,p')$ arising from these terms, we multiply by ${\cal N}^{-1/2}$ from both sides and hermitize (i.e., add Hermtian conjugate and multiply by 1/2) leading to 
\begin{eqnarray}
\bar{\cal H}^{\rm eev}_{\nu\nu'}(p,p')&=& E^{(A)}_{\nu}\;\delta_{\nu\nu'}\;\frac{\delta(p-p')}{pp'}-E^{(A)}_{\nu}\;\delta_{\nu\nu'}\sum\limits_{n=0}^{n_{\rm max}}\;\tilde R_{n\ell}(p)\;\tilde R_{n\ell}(p'),
\label{eq:a1.7}
\end{eqnarray}
for the energy eigenvalue term. Although the Hamiltonian kernel $\bar{\cal H}_{\nu\nu'}(p,p')$ in the full space (i.e. $n_{\rm max}\rightarrow \infty$) is Hermitian, the representation at finite $n_{\rm max}$ does not fulfill this symmetry and therefore necessitates the explicit hermitization. Additionally, the kinetic energy term gives
\begin{eqnarray}
\bar{\cal H}^{\rm kin}_{\nu\nu'}(p,p') 
&=& \frac{p^2}{2\mu}\;\delta_{\nu\nu'}\;\frac{\delta(p-p')}{pp'}
-\frac{p^2}{2\mu}\;\delta_{\nu\nu'}\sum\limits_{n=0}^{n_{\rm max}}\sum\limits_{n'=0}^{n_{\rm max}}
\tilde R_{n\ell}(p)\;T_{\ell,nn'}\;\tilde R_{n\ell}(p')\cr\cr
&-&\sum\limits_{n=0}^{n_{\rm max}}\;\tilde R_{n_{\rm max}+1,l}(p)\;T_{\ell,n_{\rm max}+1,n_{\rm max}}\left[ \delta_{\nu\nu'}\;\delta_{n n_{\rm max}}-\frac{1}{2} \Lambda^{-1/2}_{n\nu,n_{\rm max}\nu'}-\frac{1}{2} \Lambda^{1/2}_{n\nu, n_{\rm max}\nu'}\right]\tilde R_{n\ell}(p)\cr\cr
&-&\sum\limits_{n=0}^{n_{\rm max}}\;\tilde R_{n\ell'}(p')\left[ \delta_{\nu\nu'}\;\delta_{n n_{\rm max}}-\frac{1}{2} \Lambda^{-1/2}_{n_{\rm max}\nu, n\nu'}-\frac{1}{2} \Lambda^{1/2}_{n_{\rm max}\nu, n\nu'}\right] T_{\ell,n_{\rm max},n_{\rm max}+1}\tilde R_{n_{\rm max}+1,l}(p),
\label{eq:a1.8}
\end{eqnarray}
where $T_{\ell,nn'}\equiv \langle R_{n\ell}|\;p^2/2\mu\;|R_{n'\ell} \rangle$ are the kinetic energy matrix elements in the HO basis. The first term in Eq.~(\ref{eq:a1.7}) and Eq.~(\ref{eq:a1.8}) is respectively the
energy eigenvalue of the target nucleus and the relative nucleon-target kinetic energy. The remaining terms
constitute the respective contributions to the effective nucleon-target potential ${\cal W}_{\nu\nu'}^{\rm eev}(p,p')$ and ${\cal W}_{\nu\nu'}^{\rm kin}(p,p')$, such that
\begin{eqnarray}
\bar{\cal H}^{\rm eev}_{\nu\nu'}(p,p')&=& E^{(A)}_{\nu}\;\delta_{\nu\nu'}\;\frac{\delta(p-p')}{pp'}+{\cal W}_{\nu\nu'}^{\rm eev}(p,p'),
\label{eq:a1.9}
\end{eqnarray}
and
\begin{eqnarray}
\bar{\cal H}^{\rm kin}_{\nu\nu'}(p,p') 
&=& \frac{p^2}{2\mu}\;\delta_{\nu\nu'}\;\frac{\delta(p-p')}{pp'}+{\cal W}_{\nu\nu'}^{\rm kin}(p,p').
\label{eq:a1.10}
\end{eqnarray}
Additionally, the correction due to the average Coulomb potential is given by 
\begin{eqnarray}
\bar{\cal H}^{\rm coul}_{\nu\nu'}(p,p') 
&=& \bar V^{\rm c}_{\ell}(p,p')\;\delta_{\nu\nu'}
+\sum\limits_{n=0}^{n_{\rm max}}\sum\limits_{n'=0}^{n_{\rm max}}\left(\tilde R_{n\ell}(p)\;\langle 
R_{n'\ell'}|\;\bar V^{\rm c}_{\ell}\;|p'\rangle+\langle p|\;\bar V^{\rm c}_{\ell}\; |R_{n\ell}\rangle \tilde R_{n'\ell'}(p') \right)\left(\frac{1}{2}\;\Lambda^{-1/2}_{n\nu, n'\nu'}-\delta_{\nu\nu'}\;\delta_{n n'}\right) \cr\cr\cr
&+&\sum\limits_{n=0}^{n_{\rm max}}\sum\limits_{n'=0}^{n_{\rm max}} R_{n\ell}(p)\;R_{n'\ell'}(p')\left(\delta_{\nu\nu'}\;\tilde{\bar V}^{\rm c}_{\ell,n n'}-\frac{1}{2}\sum\limits_{n''=0}^{n_{\rm max}}\left(\tilde{\bar V}^{\rm c}_{\ell,n n''}\Lambda^{-1/2}_{n''\nu, n'\nu'}+\Lambda^{-1/2}_{n\nu, n''\nu'}\tilde{\bar V}^{\rm c}_{\ell,n'' n'}\right)\right),
\label{eq:a1.11}
\end{eqnarray}
where $\tilde{\bar V}^{\rm c}_{\ell,n n''}\equiv \langle R_{n\ell}|\bar V^{\rm c}_{\ell}|R_{n''\ell} \rangle$ are the matrix elements of average Coulomb potential in the HO basis. The full nucleon-target potential is therefore given by the sum
\begin{eqnarray}
{\cal W}_{\nu\nu'}(p,p')&=& \bar{\cal H}_{\nu\nu'}^{\rm mod}(p,p')+{\cal W}_{\nu\nu'}^{\rm eev}(p,p')+{\cal W}_{\nu\nu'}^{\rm kin}(p,p')+\bar{\cal H}^{\rm coul}_{\nu\nu'}(p,p'),
\label{eq:a1.12}
\end{eqnarray}
and the full Hamiltonian kernel takes the form
\begin{eqnarray}
\bar{\cal H}_{\nu\nu'}(p,p')&=& \left[E^{(A)}_{\nu}+\frac{p^2}{2\mu}\right]\delta_{\nu\nu'}\;\frac{\delta(p-p')}{pp'}+{\cal W}_{\nu\nu'}(p,p'),
\label{eq:a1.13}
\end{eqnarray}
  which is substituted into Eq.~(\ref{eq:a1.6}) to arrive at the effective two-body Schr\"odinger in momentum space. Finally, we can write Eq.~(\ref{eq:a1.6}) explicitly leading to
\begin{eqnarray}
\Bigg[E_{A+1}-E_{\nu}^{(A)}-\epsilon_{\nu}-\frac{p^2}{2\mu}\Bigg]\; \chi_{\nu\nu_0} (p,p_0)=\sum\limits_{\nu'}\int dp' p'^2\; {\cal W}_{\nu\nu'}(p,p')\; \chi_{\nu'\nu_0} (p',p_0).
\label{eq:a1.14}
\end{eqnarray}
We note that the quantity $E_{A+1}-\bar{\cal H}$ contains the difference $E_{A+1}-E_{\nu}^{(A)}$, which is the total energy of the ($A+1$)-system relative to the energy of the target nucleus. It is customary to define an incident relative kinetic energy $E_{\rm kin}\equiv E_{A+1}-E^{\rm (A)}_{\nu_0}$ so that the relative kinetic energy has the form energy $E_{A+1}-E_{\nu}^{(A)}=E_{\rm kin}-\epsilon_{\nu}$. The corresponding $t$~matrix fulfills  the Lippmann-Schwinger (LS) equation 
\begin{eqnarray}
t_{\nu\nu_0}(p,p_0;E_{\rm kin})=  {\cal W}_{\nu\nu_0}(p,p_0)+\sum\limits_{\nu'} \; \int dp'\;p'^2\;{\cal W}_{\nu\nu'}(p,p')\;G_{0\nu'}(E_{\rm kin},p')\;t_{\nu\nu_0}(p',p_0;E_{\rm kin}).
\label{eq:a1.15}
\end{eqnarray}  
which is equivalent to Eq.~(\ref{eq:a1.14}) complemented by the scattering boundary conditions, and the propagator is given by $G_{0\nu'}(E_{\rm kin},p')=[E_{\rm kin}-\epsilon_{\nu'}-{p'^2}/{2\mu}+i0]^{-1}$.

\bibliography{F3B.bib}
\end{document}